\documentclass[a4paper,11pt]{article}

\usepackage{amsfonts,amsmath,amssymb}
\usepackage{bm,tensor}
\usepackage{graphicx}
\usepackage{braket}
\usepackage{ifpdf}
\usepackage[utf8]{inputenc}
\usepackage{color}
\usepackage{physics}
\usepackage{jcappub}
\usepackage{mathtools}
\usepackage{subfigure}
\usepackage[normalem]{ulem}

\usepackage[colorlinks=true
,urlcolor=blue
,anchorcolor=blue
,citecolor=blue
,filecolor=blue
,linkcolor=blue
,menucolor=blue
,linktocpage=true
,pdfproducer=medialab
,pdfa=true
]{hyperref}

\title{
Gravitational waves detectable in laser interferometers from axion-SU(2) inflation}

\author[a,b]{Tomohiro Fujita,}
\author[c]{Kaname Imagawa,}
\author[c]{and Kai Murai}

\affiliation[a]{Waseda Institute for Advanced Study, Waseda University, 3-4-1 Okubo, Shinjuku, Tokyo 169-8555, Japan}
\affiliation[b]{Research Center for the Early Universe, The University of Tokyo, Bunkyo, Tokyo 113-0033, Japan}
\affiliation[c]{ICRR, The University of Tokyo, Kashiwa, 277-8582, Japan}

\emailAdd{tomofuji@aoni.waseda.jp}
\emailAdd{a0142165@icrr.u-tokyo.ac.jp}
\emailAdd{kmurai@icrr.u-tokyo.ac.jp}

\abstract{
    Chromo-natural inflation~(CNI) is an inflationary model where an axion coupled with SU$(2)$ gauge fields acts as the inflaton.
    In CNI, the gauge fields have nonzero vacuum expectation values~(VEVs), which results in the enhancement of gravitational waves~(GWs).
    The original CNI is ruled out by the Planck observations due to the overproduction of GWs.
    In this work, we consider an inflationary model where the gauge fields acquire nonzero VEVs after the CMB modes exit the horizon.
    Moreover, we add to the model another field that dominates the universe and drives inflation after the axion starts to oscillate and the gauge field VEVs vanish.
    By performing numerical simulations, we find a parameter space where the enhanced GWs do not violate the CMB constraints and can be detected by the future GWs observations such as BBO and ET.
    }


\begin{document}

\maketitle

\section{Introduction}

Primordial gravitational waves~(GWs) provide us useful information about the very early universe.
Some features of the primordial universe may be imprinted in the properties of primordial GWs such as the amplitude, spectral tilt, polarization, and non-Gaussianity, which we can explore through observations.
A well known channel of producing GWs is the quantum fluctuation of the GWs themselves during inflation~\cite{Grishchuk:1974ny,Starobinsky:1979ty}.
The amplitude of the quantum fluctuation is related to the Hubble parameter during inflation and then the observations of such primordial GWs provide a probe of the inflationary energy scale.
However, the quantum fluctuation is not a unique production channel of primordial GWs.

Another production mechanism of GWs was proposed in the chromo-natural inflation model~(henceforth ``CNI'')~\cite{Adshead:2012kp,Adshead:2012qe}.
Originally, the inflationary model using only a pseudo-scalar field or axion as an inflaton is proposed as natural inflation~\cite{Freese:1990rb,Adams:1992bn,Kim:2004rp}.
The flatness of the inflaton potential in this model is protected from large quantum corrections because of the shift symmetry.
Natural inflation is consistent with Planck observations with a super-Planckian axion decay constant~$ f \gtrsim M_{ \mathrm{Pl} } $~\cite{Freese:2004un,Savage:2006tr,Freese:2014nla}, which is inconsistent with the theory with respect to axion~\cite{Kallosh:1995hi,Banks:2003sx}.
On the other hand, CNI contains a pseudo-scalar field $\chi$ and SU(2) gauge fields with the interaction~$\chi F ^{a} _{ \mu \nu } \widetilde{F} ^{ a \mu \nu } $, which causes an effective friction for the axion.
As a result, it can take enough e-foldings to explain the observations even when~$ f \lesssim M_{ \mathrm{Pl} } $.

In the model of CNI, the gauge fields have
an isotropic and attractor background solution~\cite{Maleknejad:2013npa,Wolfson:2020fqz,Wolfson:2021fya}, which is induced by the motion of the axion.
The nonzero vacuum expectation values (VEVs) of the gauge fields induce tachyonic instabilities in the gauge field perturbations.
Moreover, the gauge field VEVs induce linear couplings between the gauge field perturbations and metric perturbations.
Thus, the enhanced gauge field perturbations linearly sources GWs, which can be much larger than those originated from vacuum fluctuations.
In this paper, we call such a system of the axion and gauge fields ``CN system''.
The model where the axion and gauge fields are dominant components in the inflationary universe and cause inflation, which we call minimal CNI, are ruled out from Planck observation~\cite{Adshead:2013qp,Adshead:2013nka}.
However, in the situation where the axion and the gauge fields are subdominant, GWs that can be detected in future GW observations can be predicted without violating the current observational constraints~\cite{Dimastrogiovanni:2016fuu,Kakizaki:2021mgj,Ishiwata:2021yne}.
In particular, the gravitational waves enhanced in the CN system are scale-dependent, chiral, and highly non-Gaussian~\cite{Dimastrogiovanni:2012ew,Maleknejad:2012fw,Maleknejad:2016qjz,Caldwell:2017chz,Agrawal:2017awz,Agrawal:2018mrg,Dimastrogiovanni:2018xnn,Fujita:2018vmv,Fujita:2018ndp,Fujita:2021flu}, which is in contrast to the non-chiral and Gaussian GWs originated from the vacuum fluctuations.
Thus, the CN system is a fascinating mechanism generating observable stochastic GWs.
Since the enhancement of GWs takes place around the horizon exit in the CN system, the frequency of the enhanced GWs corresponds to the period when the gauge fields have nonzero VEVs.
Especially, if the CN system arises in a limited interval after the CMB scale exits the horizon, the enhancement of GWs can occur only on smaller scales.
In this case, even if the axion and gauge fields dominate the universe and drive inflation, the detectable GWs can be generated remaining fluctuations on the CMB scale consistent with the observations.
Such a delayed emergence of the CN system is studied in some extensions of the original CNI~\cite{Obata:2014loa,Obata:2016tmo,Domcke:2018rvv}.

In this paper, we consider an extension of the minimal CNI, which includes two new components.
One is the phase transition of the gauge field VEVs.
The background solution of the gauge fields depends on the axion velocity, and the nonzero VEVs can appear only when the axion velocity is large enough.
Then, the gauge field can show a transition in the course of inflation depending on the shape of the axion potential.
If the CN system arises after the CMB scale exits the horizon, we can safely obtain the detectable GWs for the aforementioned reason.
The other component is multifield inflation.
If an additional field drives inflation after the axion starts to oscillate and the CN system disappears, the enhancement of the GWs does not occur at the end of the entire inflationary period.
Then, the peak frequency of the enhanced GWs can take various values depending on the duration of the period when the additional field drives inflation.
We numerically evaluate the background dynamics and then the enhancement of the GWs in this model to show that this model can predict the generation of GWs detectable by future GW experiments.

This paper is organized as follows.
In Sec.~\ref{Background Evolution}, after reviewing CNI, we explain our model and its background evolution.
In Sec.~\ref{Tensor Perturbation}, we discuss the dynamics of tensor perturbations in our model and perform the numerical simulations to evaluate the GW today.
Sec.~\ref{Summary and Discussion} is devoted to summary and discussion.
All cosmological parameters in this paper are extracted from the results of Planck 2018~\cite{Planck:2018vyg}.

\section{Background Evolution}
\label{Background Evolution}

In this section, after reviewing the CNI in Sec.~\ref{Review of Chromo-Natural Inflation}, we explain our model in Sec.~\ref{Our Model}.
In Sec.~\ref{Background Evolution sub}, we perform the numerical simulations to investigate the background evolution in our model, which will be used to solve the equations of motion for the tensor perturbations in Sec.~\ref{Tensor Perturbation}. 

\subsection{Review of Chromo-Natural Inflation~(CNI)}
\label{Review of Chromo-Natural Inflation}

In this section, we review the background dynamics of the CNI model whose action is 
\begin{align}
    S 
    =
    \int \mathrm{d} ^4 x \, \sqrt{-\tilde{g}} \, 
    [
        \mathcal{L} _{ \mathrm{EH} } 
        + \mathcal{L} _{\chi, \mathrm{gauge}} 
    ] ,
\end{align}
with
\begin{eqnarray}
    \mathcal{L} _{ \mathrm{EH} }
    &=&
    \frac{1}{2} M_{ \mathrm{Pl} } ^2 R ,
    \label{Lagrangian_EH}
    \\
    \mathcal{L} _{\chi, \mathrm{gauge}}
    &=&
    - \frac{1}{2} ( \partial \chi )^2 - V( \chi ) - \frac{1}{4} F^a _{ \mu \nu } F^{ a \mu \nu } + \frac{ \lambda \chi }{f} F^a _{ \mu \nu } \widetilde{F} ^{ a \mu \nu } ,
    \label{Lagrangian_axion_gauge}
\end{eqnarray}
where $\tilde{g}$ is the determinant of the spacetime metric,
$R$ is the Ricci scalar,
$ M_{ \mathrm{Pl} } = 1 / \sqrt{ 8 \pi G } $ is the reduced Planck mass,
$\chi$ is the axion field, $V(\chi)$ is the axion potential,
$ F^a _{ \mu \nu } \equiv \partial_{\mu} A_{\nu}^{a}-\partial_{\nu} A_{\mu}^{a}-g \epsilon^{a b c} A_{\mu}^{b} A_{\nu}^{c} $ is the field strength of the SU$(2)$ gauge fields~$ A^a _{\mu} $,
~$g$ is the gauge coupling,
$\widetilde{F}^{a \mu \nu} \equiv \epsilon^{\mu \nu \rho \sigma} F_{\rho \sigma}^{a} /(2 \sqrt{-\tilde{g}})$ is the dual of $F^a _{ \mu \nu }$,
$\lambda$ is the Chern--Simons coupling,
and~$f$ is the axion decay constant.
We consider the following ansatz for the background components of the gauge fields,
\begin{equation}
    A_{0}^{a}
    =
    0,
    \quad
    A_{i}^{a}
    =
    \delta_{i}^{a} a(t) Q(t) ,
    \label{gauge_field_ansatz}
\end{equation}
which is an attractor solution and respects the homogeneity and isotropy~\cite{Adshead:2012kp,Maleknejad:2013npa,Wolfson:2020fqz,Wolfson:2021fya}.

\begin{figure}[t]
    \begin{center}
        \includegraphics[clip,width=13.5cm]{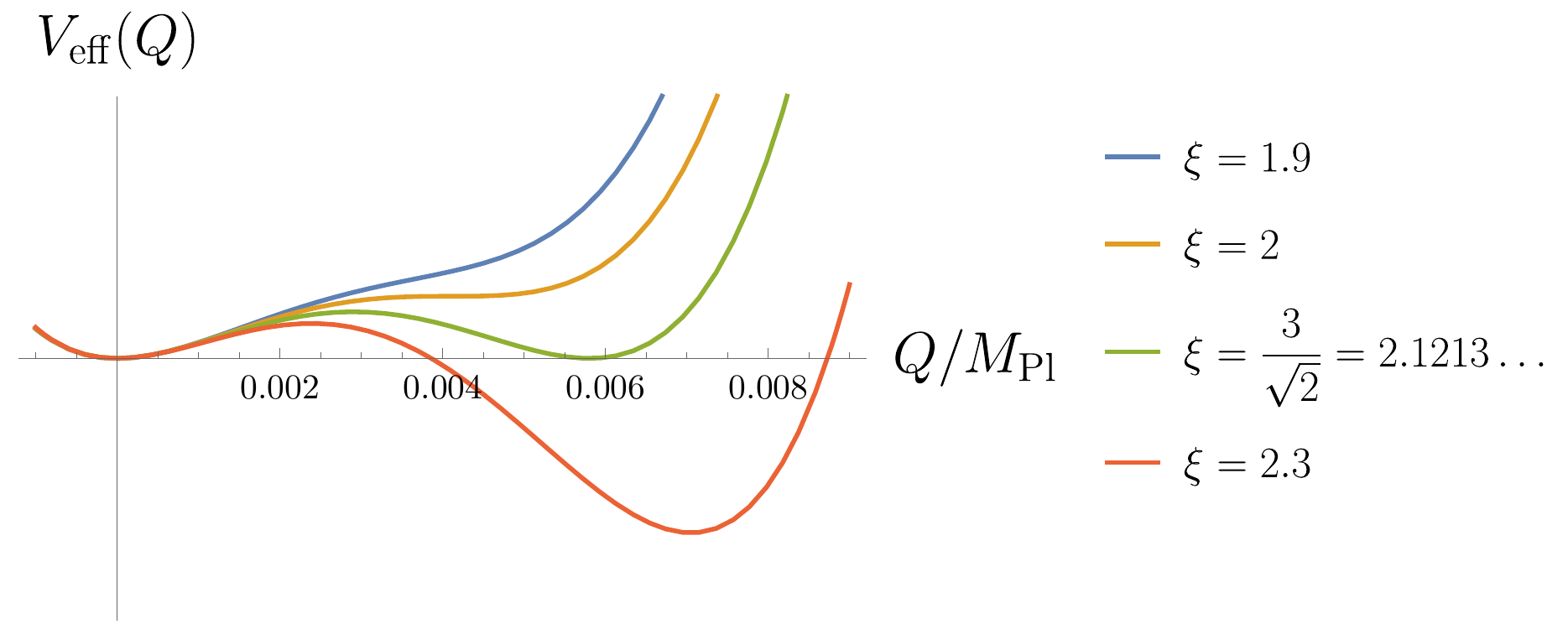}
        \caption{
        Effective potential of the background gauge field against $Q/M_\mathrm{Pl}$ with $( g, H ) = ( 10^{-3}, 10^{13}~\mathrm{GeV} )$.
        The colored lines denote the cases with $\xi=1.9$ (blue), $2.0$ (orange), $\xi_\mathrm{cr}$ (green) and $2.3$ (red). 
        The critical value $ \xi _{ \mathrm{cr} } \approx 2.12$ is defined in Eq.~\eqref{critial_value_xi}.
        When $\xi< \xi _{ \mathrm{cr} } $, the true vacuum is located at the origin~($ Q = 0 $). When $\xi>\xi_{ \mathrm{cr} } $, the true vacuum resides at a nonzero value~($ Q =Q_+ \neq 0 $).
        }
        \label{Potential_guage_field}
    \end{center}
\end{figure}

The Friedmann equation reads
\begin{equation}
    3 M_{ \mathrm{Pl} }^2 H^2
    =
    \rho _{ \chi } + \rho _{ \mathrm{E} } + \rho _{ \mathrm{B} } ,
\end{equation}
where $\rho _{\chi}$, $\rho _{ \mathrm{E} }$, and $\rho _{ \mathrm{B} }$ are the energy densities of the axion, electric, and magnetic fields, respectively, which are defined by
\begin{equation}
    \rho _{ \chi }
    =
    \frac{1}{2} \dot{\chi} ^2 + V( \chi ) ,
    \label{energy_density_axion}
    \qquad
    \rho _{ \mathrm{E} }
    =
    \frac{3}{2} \left( \dot{Q} + HQ \right)^2 ,
    \qquad
    \rho _{ \mathrm{B} }
    =
    \frac{3}{2} g^2 Q^4 .
\end{equation}
Here, the dot denotes the derivative with respect to the physical time $t$.
The equations of motion for the axion and gauge fields read
\begin{eqnarray}
    &&
    \ddot{\chi} + 3 H \dot{\chi} + V ^{\prime}( \chi )
    =
    - \frac{ 3 g \lambda }{ f } Q^2 \left( \dot{Q} + H Q \right) ,
    \label{Background_EoM_axion}
    \\
    &&
    \ddot{Q} + 3 H \dot{Q} + \left( 2 H^2 + \dot{H} \right) Q + 2 g^2 Q^3
    =
    \frac{ g \lambda }{ f } \dot{\chi} Q^2 ,
    \label{Background_EoM_gauge_field}
\end{eqnarray}
where the prime denotes the partial derivative with respect to~$\chi$.
Eq.~\eqref{Background_EoM_gauge_field} can be rewritten as
\begin{equation}
    \ddot{Q} + 3 H \dot{Q} + \frac{ \partial V_{ \mathrm{eff} } (Q) }{ \partial Q }
    =
    0 ,
\end{equation}
where $V_{ \mathrm{eff} } (Q)$ is the effective potential of the gauge field,
\begin{equation}
    V_{ \mathrm{eff} } (Q)
    =
    \frac{1}{2} \left( 2 H^2 + \dot{H} \right) Q^2 - \frac{ 2 g \xi }{3} H Q^3 + \frac{ g^2 }{2} Q^4.
    \label{effective_potential_gauge_field}
\end{equation}
Here $\xi \equiv \lambda \dot{\chi}/(2 f H)$ is a dimensionless parameter proportional to the axion velocity.
In this subsection, we treat~$\xi$ as a constant parameter and take the slow-roll limit $\dot{H} = 0$.
Fig.~\ref{Potential_guage_field} presents this effective potential for various values of $\xi$.
When~$ \xi > 2 $, this potential has local minima at~$ Q = 0 $, $Q_{+}$ and a local maximum at~$Q = Q_{-}$ satisfying
\begin{eqnarray}
    \frac{ \partial V_{ \mathrm{eff} } (Q) }{ \partial Q }
    =
    0
    \, \, \, 
    & \quad\Leftrightarrow\quad &
    \, \, \, 
    Q
    =
    0
    ,
    \, 
    Q_{\pm},
    \label{Q_local}
\end{eqnarray}
with
\begin{equation}
    Q_\pm \equiv \frac{ \xi \pm \sqrt{ \xi ^2 -4 } }{ 2 g } \, H.
\end{equation}
Note that $ Q = Q_{ \pm } $ is equivalent to
\begin{equation}
    \xi
    =
    m_Q + m_Q ^{-1} ,
    \label{ximQ}
\end{equation}
where~$ m_Q \equiv g Q / H $ is the effective mass of the gauge fields normalized by the Hubble parameter.
Assuming $\dot{H}=0$, one finds that $ g^2 V_{ \mathrm{eff} }(Q) / H^4 $ contains only~$\xi$ and~$m_Q$.
Thus, $\xi$ determines the vacuum structure of the background gauge fields. The critical value of $\xi$ is given by
\begin{equation}
    V_\mathrm{eff}(Q_+)
    =
    0
    \quad \Leftrightarrow \quad
    \xi _{ \mathrm{cr} }
    =
    \frac{3}{\sqrt{2}}
    \approx
    2.121.
    \label{critial_value_xi}
\end{equation}
As seen in Fig.~\ref{Potential_guage_field}, when $ \xi < \xi _{ \mathrm{cr} } $, the true vacuum is located at the origin and the gauge fields are expected to have no VEV.
Indeed, the trivial configuration $Q(t) = 0$ is a solution of Eq.~\eqref{Background_EoM_gauge_field}.
Especially for sufficiently small $\xi$, the axion field dynamics becomes virtually the same as that in a single-field slow-roll inflation.
On the other hand, when $ \xi > \xi _{ \mathrm{cr} } $, the gauge fields have nonzero VEVs at the true vacuum, which causes a non-trivial dynamics distinct from the single-field slow-roll case.

If $\xi$ is smaller than $\xi_\mathrm{cr}$ at first and then increases in time, it may exceed the critical value at some point during inflation.
In this case, we expect that a phase transition of the background gauge fields occurs.
Even if $Q(t)$ originally vanishes, it acquires a nonzero VEV and is stabilized at $Q=Q_+$ after $\xi$ exceeds $\xi_\mathrm{cr}$.
We employ this phase transition in our model as discussed in what follows.

\subsection{Our Model}
\label{Our Model}

Now, we explain our model.
The action is given by
\begin{equation}
    S
    =
    \int \mathrm{d} ^4 x \, 
    \sqrt{- \tilde{g}}
    [ \mathcal{L} _{ \mathrm{EH} } + \mathcal{L} _{ \chi , \mathrm{gauge} } + \mathcal{L} _{ \mathrm{add} } ] ,
\end{equation}
where~$\mathcal{L} _{ \mathrm{EH} }$ and~$ \mathcal{L} _{ \chi , \mathrm{gauge} } $ are the same as Eqs.~\eqref{Lagrangian_EH} and \eqref{Lagrangian_axion_gauge}, while~$\mathcal{L} _{ \mathrm{add} }$ is the action for an additional inflaton field, which induces the third inflationary stage as we will explain below.
We assume that the additional field does not couple with the axion or gauge fields.
Since the dynamics of the additional field hardly affects the GW enhancement, we do not specify the action of the additional field.

The Friedmann equation reads
\begin{equation}
    3 M_{ \mathrm{Pl} } ^2 H^2
    =
    \rho _{\chi} + \rho _{ \mathrm{E} } + \rho _{ \mathrm{B} } + \rho _{ \mathrm{add} } ,
    \label{Friedmann_eq_our_model}
\end{equation}
where the background gauge fields, if any, are assumed to take the ansatz~\eqref{gauge_field_ansatz}.
$\rho _{\chi}$, $\rho _{ \mathrm{E} }$, and~$\rho _{ \mathrm{B} }$ are the same as Eq.~\eqref{energy_density_axion}, and~$\rho _{ \mathrm{add} }$ denotes the energy density of the additional field. 
The equations of motion for the axion and gauge field are also the same as Eqs.~\eqref{Background_EoM_axion} and~\eqref{Background_EoM_gauge_field}.
For later convenience, we also introduce the slow-roll parameters as
\begin{equation}
    \epsilon
    \equiv
    - \frac{ \dot{H} }{H^2}
    =
    \epsilon _{ \chi }  + \epsilon _{ \mathrm{B} } + \epsilon _{ \mathrm{E} }
    + \epsilon_\mathrm{add},
    \label{epsilon_slow_roll}
    \qquad
    \eta
    \equiv
    \frac{ \dot{\epsilon} }{ H \epsilon } ,
\end{equation}
with
\begin{equation}
    \epsilon _{ \chi }
    =
    \frac{ \dot{\chi} ^2 }{ 2 H^2 M_{ \mathrm{Pl} } ^2 } ,
    \qquad
    \epsilon _{ \mathrm{E} }
    =
    \frac{ \left(  H Q + \dot{Q} \right) ^2 }{ H^2 M_{ \mathrm{Pl} } ^2 } , 
    \qquad
    \epsilon _{ \mathrm{B} }
    =
    \frac{ g^2 Q^4 }{ H^2 M_{ \mathrm{Pl} } ^2 } .
    \label{slow_roll_B}
\end{equation}
Here, $\epsilon_\mathrm{add}$ denotes the slow-roll parameter of the additional field.
In the following analysis, we approximate $\rho_\mathrm{add}$ to be constant, and thus we neglect $\epsilon_\mathrm{add}$. 

We adopt the axion monodromy potential for the axion potential~\cite{Silverstein:2008sg,McAllister:2008hb}:
\begin{equation}
    V( \chi )
    =
    \mu ^4 \left[ \left( 1 + \frac{ \chi ^2 }{ f^2 } \right)^{ \frac{p}{2} } - 1 \right] ,
    \label{potential_axion}
\end{equation}
where $\mu$ is a dimensionful parameter which determines the inflationary scale.
Since this potential becomes sufficiently flat for small $p$, the model predictions can be consistent with the Planck observation even in a single-field slow-roll inflation case.
In this paper, we adopt the following model parameters.
\begin{equation}
    p=0.1, \qquad f=M_{ \mathrm{Pl} }, \qquad \lambda=25, \qquad g=  10^{-3}.
    \label{parameter values}
\end{equation}
The value of $\mu$ will be fixed later.
Note that, while $\lambda$ is often taken to be $ \mathcal{O} (100) $ in the conventional CNI model, we employ a smaller number.
Here, we chose the value of $p$ so that $(n_s, r)$ can be consistent with the observational constraints as we will see later in Fig.~\ref{Planck_constraint}.
    If we take larger $p$, the value of $r$ also becomes larger for the same $n_s$.

\begin{figure}[htpb]
    \begin{center}
        \includegraphics[clip,width=9cm]{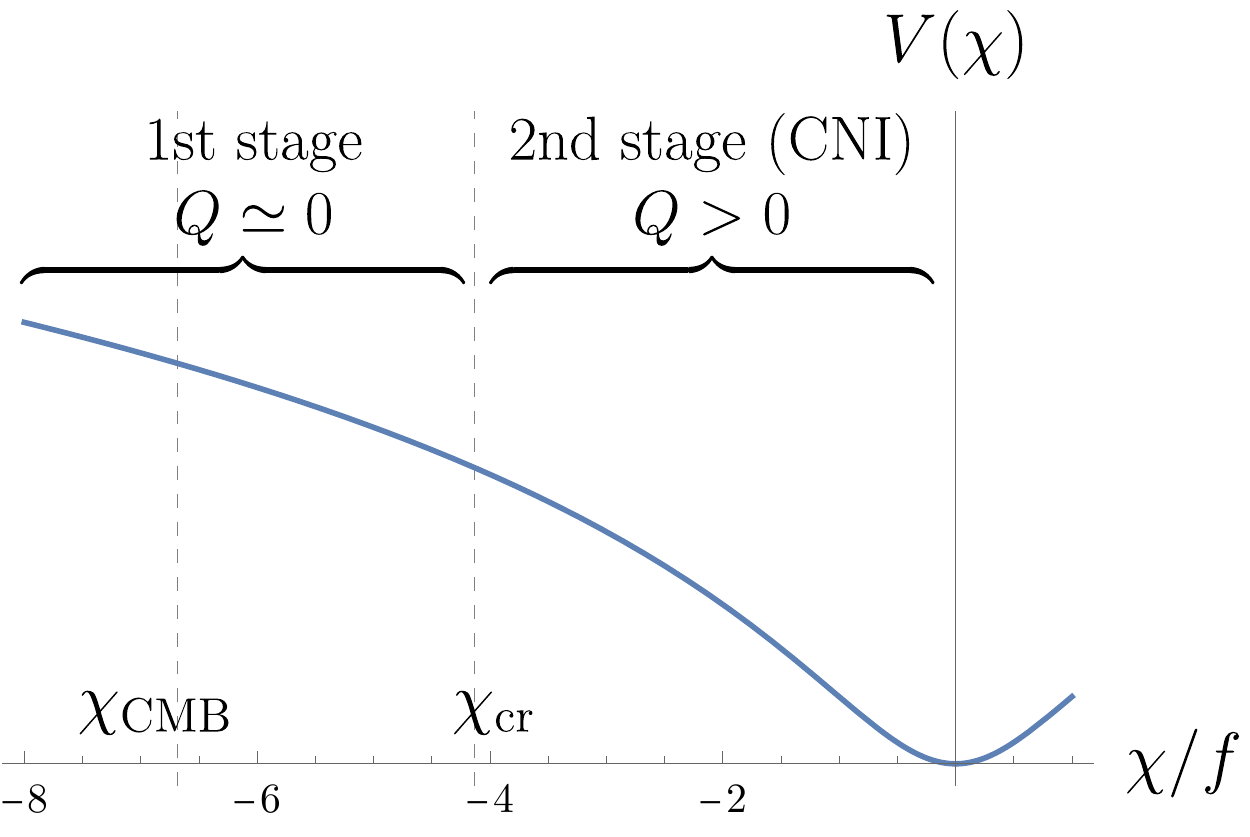}
        \caption{
            Axion potential with~$ p = 0.1 $ in Eq.~\eqref{potential_axion}. We assume that the axion initial value is negative.
            When the axion $\chi$ reaches $ \chi _{ \mathrm{CMB} } $, the CMB scale exits the horizon. 
            $ \chi _{ \mathrm{cr} } (>\chi_{ \mathrm{CMB}})$ is the axion field value at which $\xi$ exceeds the critical value $\xi_\mathrm{cr}$, and is a boundary between the first and second stages of inflation.
            The background gauge fields vanishes in the first stage $(\xi<\xi_\mathrm{cr})$, while they have nonzero VEVs during the second stage $(\xi>\xi_\mathrm{cr})$.
            When the axion begins to oscillate around its potential minimum, the second stage ends and the additional field dominates the universe initiating the third stage of inflation.
        }
        \label{Axion_potential}
    \end{center}
\end{figure}
Figure~\ref{Axion_potential} shows the shape of the axion potential and the schematic of the axion dynamics in our model.
The axionic inflaton starts to roll down its potential with a negative initial value. 
At the beginning, $\xi$ is smaller than $\xi_\mathrm{cr}$, and the background gauge fields have no VEV.
We regard this phase as a single-field slow-roll phase by neglecting the Chern--Simons coupling and call it the first stage of inflation.
Strictly speaking, the coupling between the inflaton and the gauge fields can be significant even without the VEV, but we confirm its effect is negligible during our first stage in Appendix~\ref{The effect of the gauge fields during the first stage}.
The CMB scale exits the horizon during this first stage. 
Since $|V'(\chi)|$ increases for $\chi<-f/\sqrt{1-p}$ and the axion accelerates, $\xi$ eventually exceeds the critical value $\xi_\mathrm{cr}$ and the background gauge fields acquire nonzero VEVs.
Once $Q$ is stabilized at $Q_+$, inflation turns into the CNI phase and we call it the second stage of inflation.
Finally, when the axion approaches the potential minimum and its slow-roll condition is violated, the second stage ends.
As the axion and gauge fields quickly lose their energy, the additional field dominates the total energy density and the third stage of inflation begins.
In short, $\xi=\xi_\mathrm{cr}$ triggers the transition from the first into the second stage, while the end of the CNI phase characterized by $\epsilon=1$ prompts the onset of the third stage.

Table.~\ref{Our_model_setup} summarizes our inflationary scenario with the three stages.
\begin{table}[t]
    \centering
    \begin{tabular}{c|ccc}
        & 1st stage & 2nd stage & 3rd stage \\
        \hline \hline
        \\
        Inflation
        &
        \begin{tabular}{c}
            Single field inflation \\
            of axion
        \end{tabular}
        &
        CNI
        &
        \begin{tabular}{c}
            Single field inflation \\
            of an additional field
        \end{tabular}
        \\
        \\
        $\xi$ & $\xi \lesssim \xi _{ \mathrm{cr} } $ & $\xi \gtrsim \xi _{ \mathrm{cr} } $ & Damped oscillation
        \\
        \\
        $Q$ & $Q = 0$ & $Q > 0$ & Damped oscillation
        \\
        \\
        End & $ \xi \simeq \xi _{ \mathrm{cr} } $ & $ \epsilon = 1 $ & ---
        \\
        \\
    \end{tabular}
    \caption{
        Three stages of our inflationary model.
        In the first stage, $\xi$ takes a smaller value than in the second stage because the axion potential is flatter for larger $|\chi|$.
        Then, the SU$(2)$ gauge fields have no VEV.
        We assume that the CMB scale exits the horizon during the first stage. 
        As the value of $\xi$ increases, the configuration with a nonzero $Q$ becomes the true vacuum as seen in Fig.~\ref{Potential_guage_field}.
        Then, the system of the axion and gauge fields experiences the phase transition, which occur roughly when~$\xi$ exceeds $ \xi _{ \mathrm{cr} } $.
        In the second stage of the inflation, the gauge fields have nonzero VEVs.
        As the axion field approaches the minimum of the potential, the CNI phase ends and the third stage starts. 
        We define the end of the second stage by~$ \epsilon = 1 $, which is introduced in Eq.~\eqref{epsilon_slow_roll}.
        In the third stage, an additional field dominates the universe, which is different from the axion field.
        }
    \label{Our_model_setup}
\end{table}
This model is motivated for the following reasons.
Unfortunately, the minimal model of CNI generates gravitational waves with too large amplitude on the CMB scale and fails to satisfy the observational upper bound of the tensor-to-scalar ratio for the suitable scalar tilt.
If $\xi$ is smaller than $ \xi _{ \mathrm{cr} }$ in the early stage of inflation, however, it would be free from the overproduction of the tensor modes on the CMB scale, because the gauge fields vanish $Q(t) = 0 $ and only the axion field contributes the inflationary dynamics.
As the axion inflation proceeds, $\xi$ increases and the system goes into the CNI phase. Thus, as a unique observational signature of this model, we expect the enhancement of gravitational waves in the chromo-natural system during the second stage, which may be detectable by future GW observations.

\subsection{Background Evolution}
\label{Background Evolution sub}

Our inflationary scenario consists of three stages as explained in the previous section. Here, we solve the background evolution during each stage in order.

\subsubsection{The first stage: single-field slow-roll phase}
\label{The first stage: single-field slow-roll phase}

During the first stage, inflation is driven only by the axion and we analyze this phase following the standard procedure of single-field slow-roll inflation.
The scalar tilt, tensor-to-scalar ratio and curvature perturbation are described in terms of the slow-roll parameters as
\begin{equation}
    ( n_s, \, r, \, \mathcal{P} _{\mathcal{R}} )
    =
    \left( 1 - 6 \epsilon _{\chi, \mathrm{v} } + 2 \eta _{\chi, \mathrm{v} }, \, 16 \epsilon _{\chi, \mathrm{v} }, \, \frac{ V(\chi) }{ 24 \pi ^2 \epsilon _{\chi} } \right),
    \label{ns_r_curvature}
\end{equation}
with
\begin{equation}
    \epsilon _{\chi, \mathrm{v} }
    =
    \frac{ M_{ \mathrm{Pl} }^2 }{2} \left[ \frac{ V^{\prime} (\chi) }{ V(\chi) } \right]^2 ,
    \, \, \, 
    \eta _{\chi, \mathrm{v} }
    =
    M_{ \mathrm{Pl} } ^2 \, \frac{ V^{ \prime \prime } (\chi) }{ V(\chi) } .
\end{equation}
$\xi$ is also determined by the slow-roll parameter,
\begin{equation}
    \xi
     = \frac{\lambda}{2 f H}\dot{\chi}
    \simeq
    \frac{ \lambda }{f} \, \sqrt{ \frac{ \epsilon _{ \chi, \mathrm{v} } }{2} } ,
    \label{xi_slowroll_relation}
\end{equation}
where we use~$ \epsilon _{\chi} \simeq \epsilon _{ \chi, \mathrm{v} } $.
$\chi_{\mathrm{cr}}$ can be analytically obtained through this equation with $\xi=\xi_\mathrm{cr}$.

These quantities in Eq.~\eqref{ns_r_curvature} have been constrained by the CMB observations.
To obtain the model predictions to be compared with the observational constraints, we determine $\chi_\mathrm{CMB}$ and $\mu$ as a function of one free parameter in the following way.
First, we require that the model prediction of the curvature perturbation~\eqref{ns_r_curvature} match the result of the Planck observation~\cite{Planck:2018vyg},
\begin{equation}
    \mathcal{P} _{ \mathcal{R} } ( \chi _{ \mathrm{CMB} }, \mu )
    =
    2.2 \times 10^{-9},
    \label{observation_curvature_perturbation}
\end{equation}
which leads to a relation between $\chi_\mathrm{CMB}$ and $\mu$.
Next, we consider the sum of the e-folding numbers of the first and second stages
\begin{equation}
    \Delta N
    \equiv
    N_{ \mathrm{1st} }( \chi _{ \mathrm{CMB} } ) + N_{ \mathrm{2nd} } ( \mu ),
    \label{delta N}
\end{equation}
as a free parameter.
The e-foldings of the first inflationary stage $N_\mathrm{1st}$ is given by
\begin{equation}
    N_{ \mathrm{1st} } ( \chi _{ \mathrm{CMB} } )
    \simeq
    \int _{ \chi _{ \mathrm{cr} } } ^{ \chi _{ \mathrm{CMB} } } \, \frac{ \mathrm{d} \chi }{ \sqrt{ 2 \epsilon _{ \chi, \mathrm{v} } } },
    \label{e-folds_1st}
\end{equation}
where we used $ \epsilon _{\chi} \simeq \epsilon _{ \chi, \mathrm{v} } $ again. 
On the other hand, $N_\mathrm{2nd}$ can be obtained for various values of~$\mu$ by numerically solving the CN system, that is, Eqs.~\eqref{Background_EoM_axion} and \eqref{Background_EoM_gauge_field}.
Note that $N_\mathrm{2nd}$ does not depend on $\chi_\mathrm{CMB}$ since the second stage starts at $\chi=\chi_\mathrm{cr}$.
The details of this calculation will be explained in the next section. 
As a result, Eq.~\eqref{delta N} gives us another relation between $\chi_\mathrm{CMB}$ and $\mu$ for a fixed $\Delta N$.
By combining the two relations, we finally obtain $\chi_\mathrm{CMB}$ and $\mu$ as a function of $\Delta N$.
Substituting the obtained $ \chi _{ \mathrm{CMB} } $ and $\mu$ into Eq.~\eqref{ns_r_curvature}, we find the model predictions for the scalar tilt and tensor-to-scalar ratio.
Note that, as discussed below, the duration of the second stage of inflation can be modified for the same values of $(\chi_\mathrm{CMB}, \mu)$ and $(n_s, r)$.
Thus, in this case, $\Delta N$ is interpreted just as a label of $(\chi_\mathrm{CMB}, \mu)$ consistent with $\mathcal{P}_\zeta$.

\begin{figure}[htpb]
    \begin{center}
        \includegraphics[clip,width=9.5cm]{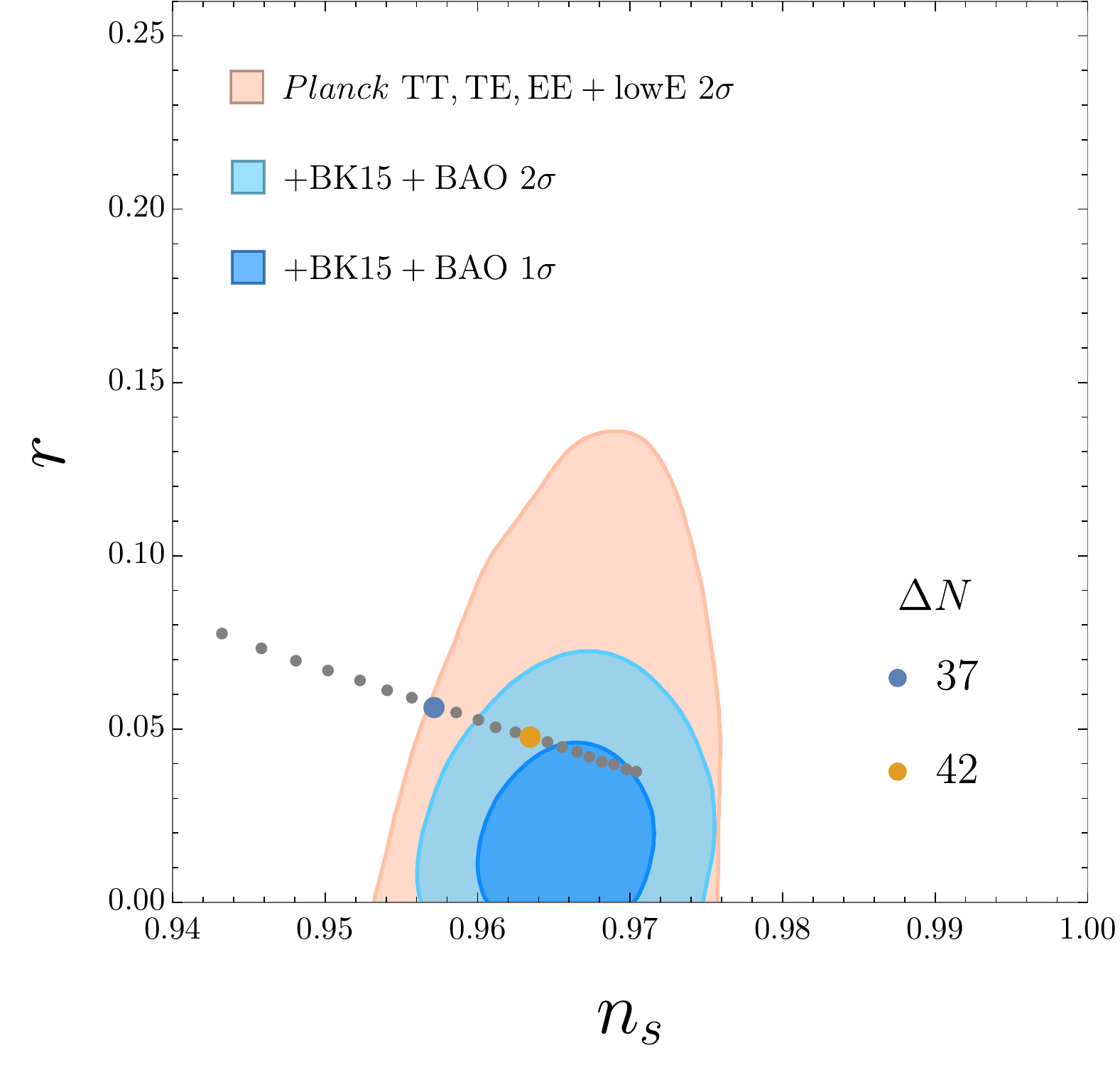}
        \caption{
            Theoretical prediction (dots) and the constraints from observations (contour) for the scalar tilt~$n_s$ and tensor-to-scalar ratio~$r$.
            The blue and orange dots denote the cases with $ \Delta N = N_{ \mathrm{1st} } + N_{ \mathrm{2nd} } = 37$ and~$42$, respectively.
            The other gray dots are shown in increments of $\Delta N$ by~$1$.
            For all dots, the parameter $\mu$ is chosen to satisfy the normalization constraint~\eqref{observation_curvature_perturbation}.
            The orange, light blue, and dark blue regions represent the constraints from different combinations of the observations~\cite{Planck:2018vyg}.
        }
        \label{Planck_constraint}
    \end{center}
\end{figure}

In Fig.~\ref{Planck_constraint}, we compare these predictions with the observational constraints.
The gray dots in Fig.~\ref{Planck_constraint} represent the theoretical prediction in our model plotted in increments of e-foldings~$\Delta N$ by~$1$ from~$30$ to~$50$.
For a sufficiently large $\Delta N$, the model predictions are compatible with the observational constraint.

In the analysis of the first stage, we ignore the coupling between the inflaton and the gauge fields. However, one may wonder if this coupling affects the inflaton's background dynamics or the curvature perturbation. In Appendix~\ref{The effect of the gauge fields during the first stage}, we confirm the effect of the coupling to the gauge fields is negligible during the first stage and verify our approximation.

\subsubsection{The second stage: CNI phase}

In the second stage, inflation is driven by the chromo-natural system. 
We can numerically solve Eqs.~\eqref{Background_EoM_axion} and \eqref{Background_EoM_gauge_field}, while we have an issue of the initial condition.
Although the axion initial value  is straightforwardly set to~$ \chi _{ \mathrm{cr} } $, that of the background gauge field is non-trivial.
Note that $Q=0$ remains a classical solution of Eq.~\eqref{Background_EoM_gauge_field} even for $\xi>\xi_\mathrm{cr}$.
The quantum fluctuations of the gauge fields are expected to initiate the transition, while our background EoM does not incorporate them.
As a simple approximation, therefore, we assume that, when~$\xi$ reaches $ \xi _{ \mathrm{cr} } $, $Q$ instantly change from~$0$ to~$Q_{-}$.

When the gauge fields have no VEV, the SU(2) gauge fields behave like three copies of U(1) gauge fields and a part of the gauge field perturbations significantly grows around the horizon exit.
After the horizon exit, such perturbations effectively contribute to the background gauge field.
Since the magnitude of the amplified perturbations exponentially depends on $\xi$, the background gauge fields are expected to grow enough to overcome the barrier of the effective potential with large $\xi$~\cite{Domcke:2018rvv}.
However, the emergence of the background gauge fields is a stochastic process and difficult to precisely evaluate.
Then, we assume that the background gauge fields acquire an amplitude large enough to settle into the nonzero background solution when $\xi$ grows up to $\xi = \xi_\mathrm{cr}$ and, at that time, we set $Q = Q_-$.

\begin{figure}
    \subfigure{%
        \includegraphics[width=7.6cm]{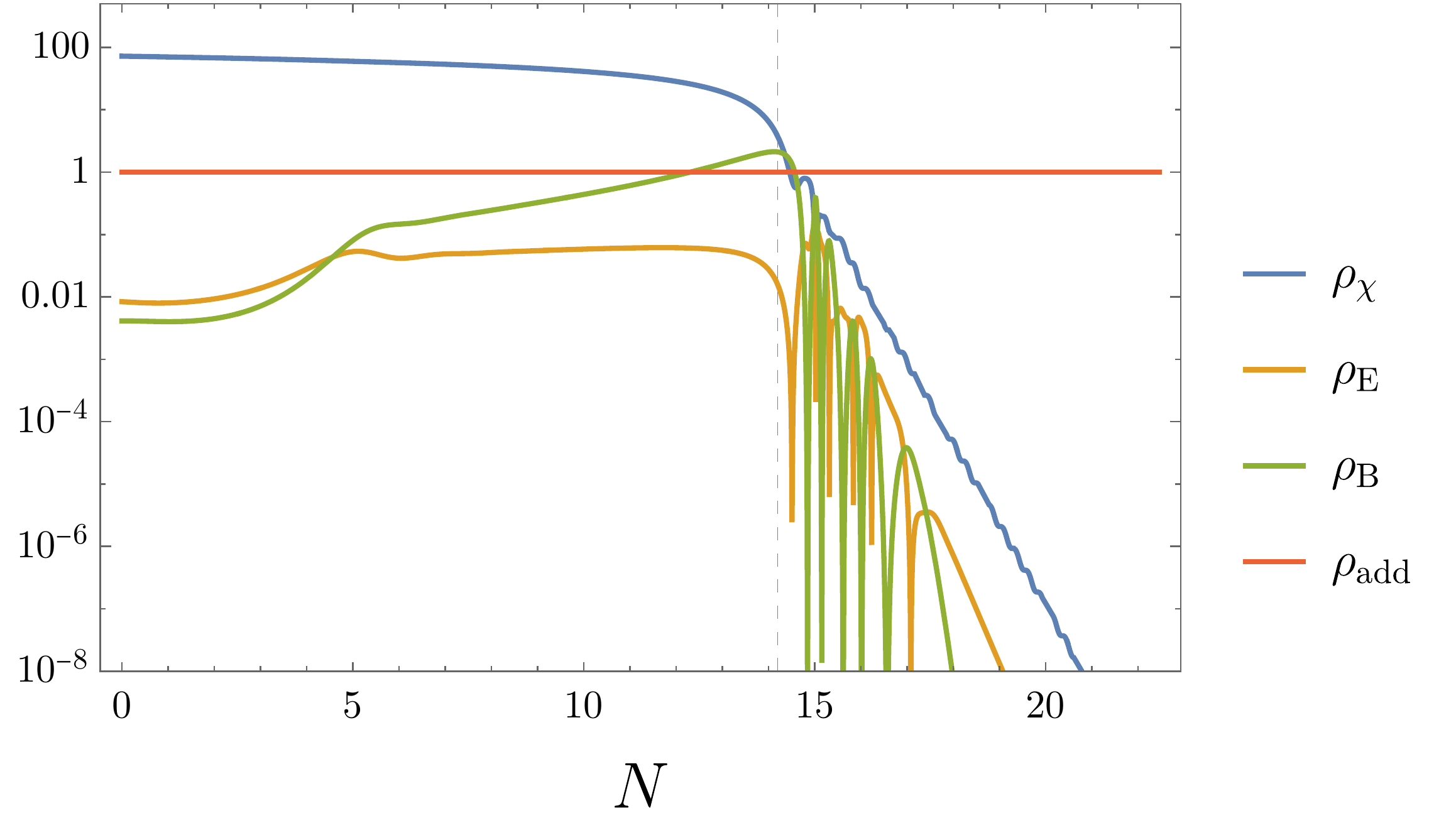}}%
    \subfigure{
        \includegraphics[width=7.8cm]{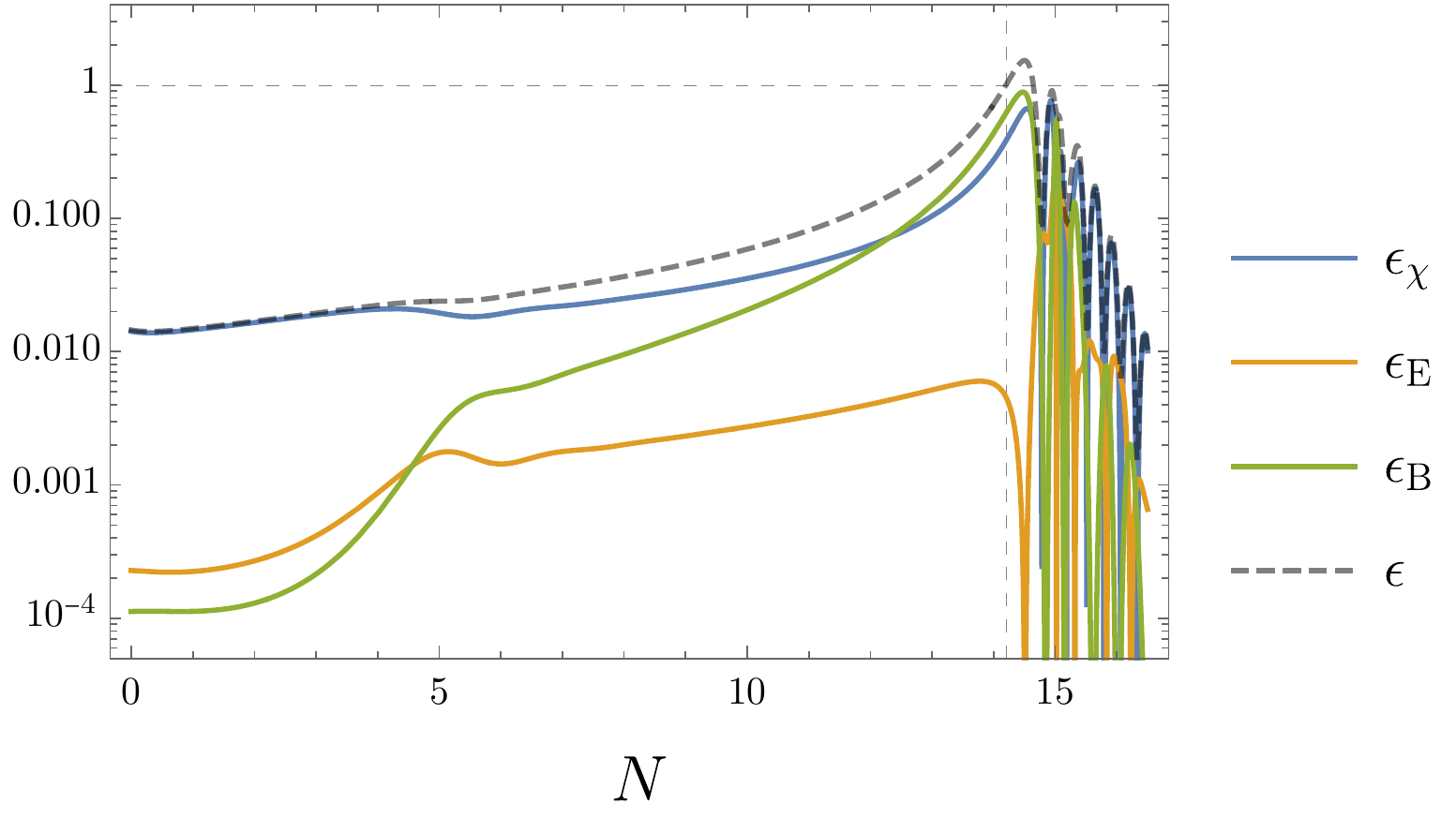}}
    \caption{
        Time evolution of the energy densities~(left panel) and slow-roll parameters~(right panel).
        The vertical dashed lines represent the end of the second inflationary stage defined by~$ \epsilon = 1 $.
        In the left panel, the blue, orange, green, and red lines represent the energy densities of the axion, electric, magnetic, and additional fields, which are defined in Eq.~\eqref{energy_density_axion}.
        These energy densities are normalized by $\rho_\mathrm{add}$ which is assumed to be constant and dominates the universe during the third stage.
        In the right panel, the blue, orange, green, and gray lines represent the slow-roll parameters for the axion, electric, magnetic, and all fields, which are defined in Eqs.~\eqref{epsilon_slow_roll} and \eqref{slow_roll_B}.
        }
    \label{evolution_of_background_parameterset2} 
\end{figure}

For simplicity, we also set the time derivative of~$Q$ to be zero at the initial time of the second stage.
We solve Eqs.~\eqref{Background_EoM_axion}, \eqref{Q_local}, and~$ \dot{\chi} = 2 f H \xi _{ \mathrm{cr} } / \lambda $ to obtain the initial values of $H$, $Q$, and the time derivative of~$\chi$.
In other words, $\chi$ and $\xi$ are continuous while $H, \dot{\chi}$, and $Q$ jump between the first and second stages.
As shown in Fig.~\ref{evolution_of_background_parameterset2}, however, the contribution of the gauge fields to the total energy is only an order of $10^{-4}$ and the discontinuities of $H$ and $\dot{\chi}$ are negligibly small.

\begin{figure}
    \subfigure{%
        \includegraphics[width=7.0cm]{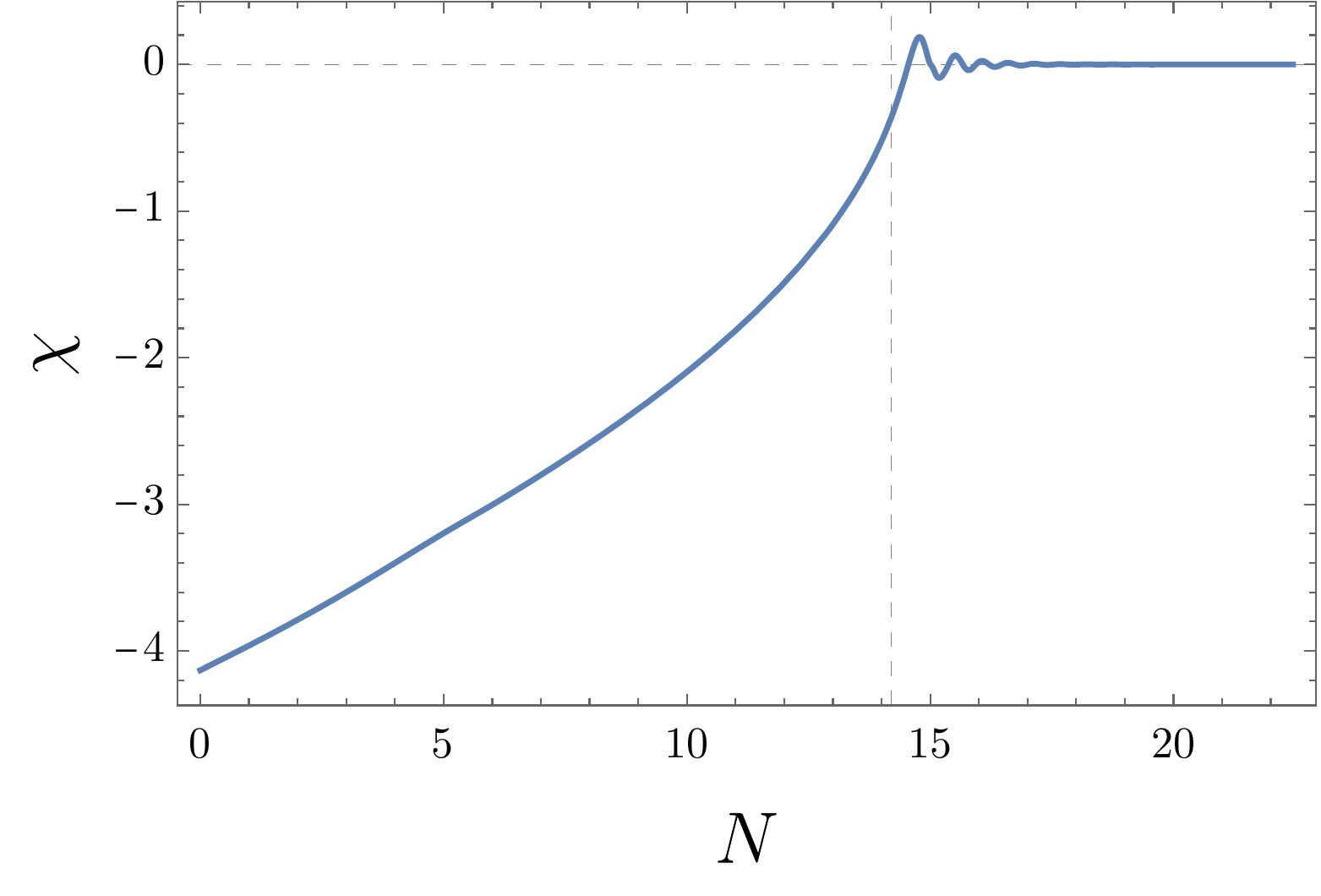}}%
    \subfigure{
        \includegraphics[width=7.3cm]{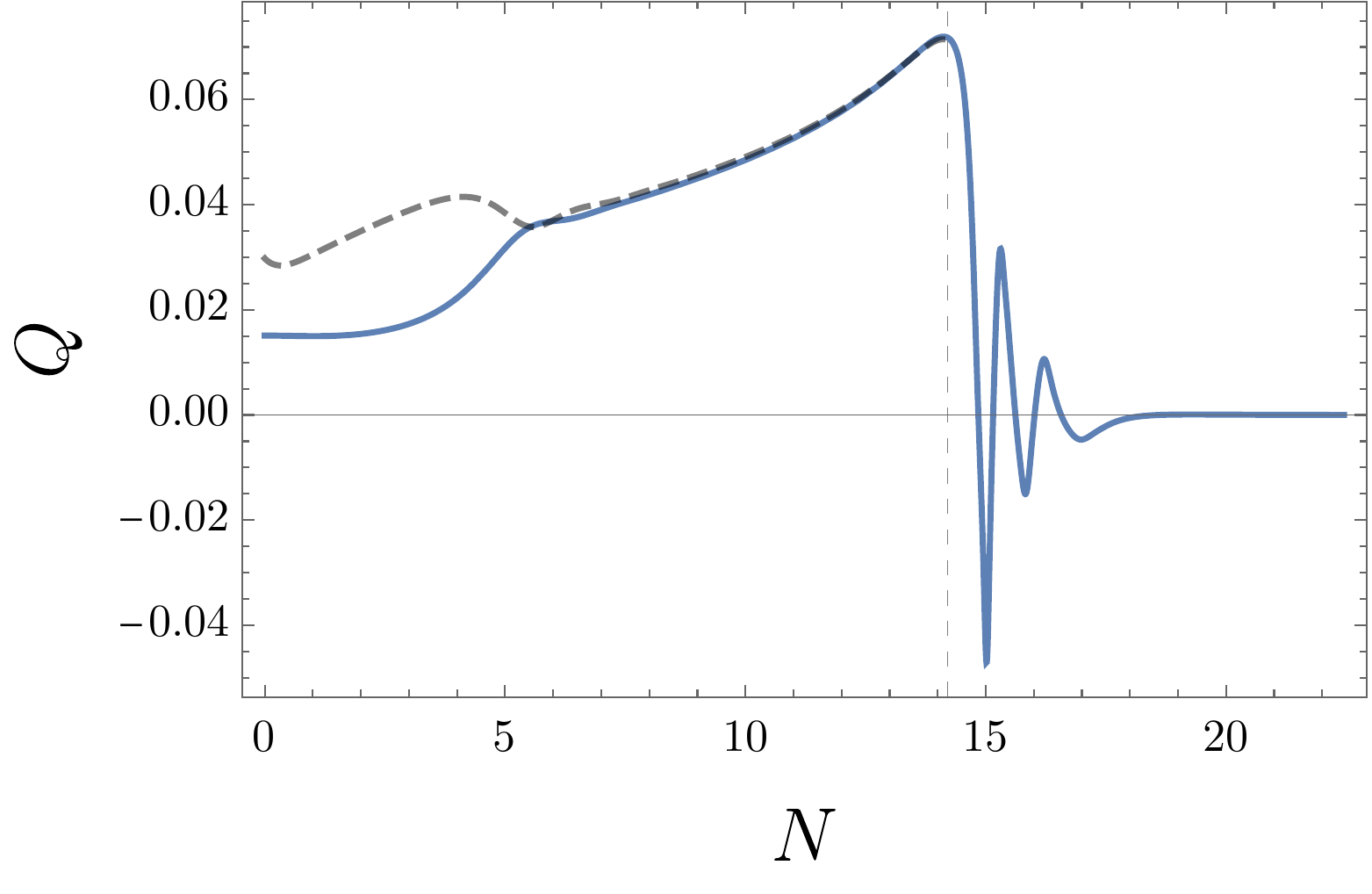}}
    \subfigure{%
        \includegraphics[width=7.0cm]{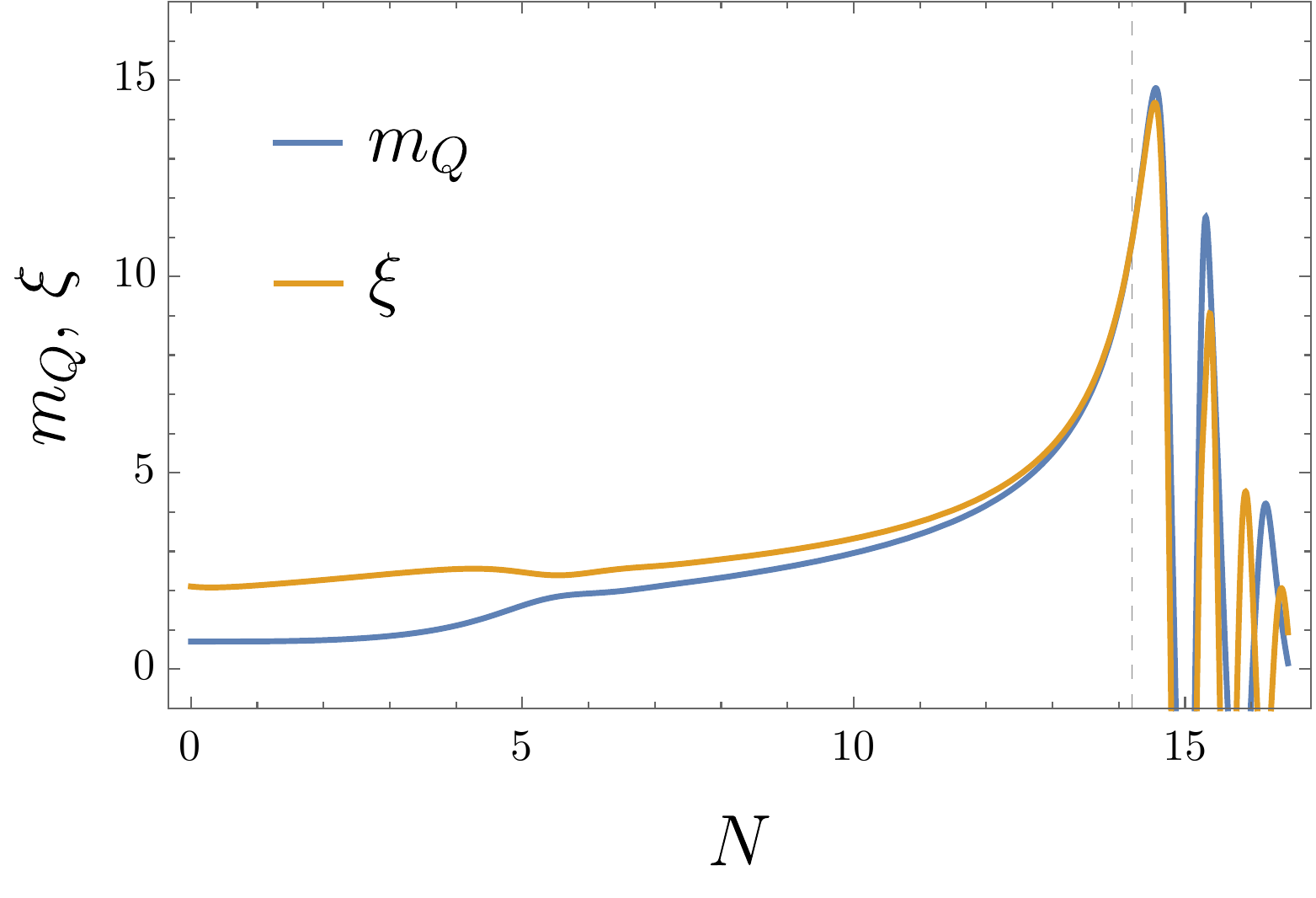}}%
    \subfigure{
        \includegraphics[width=7.6cm]{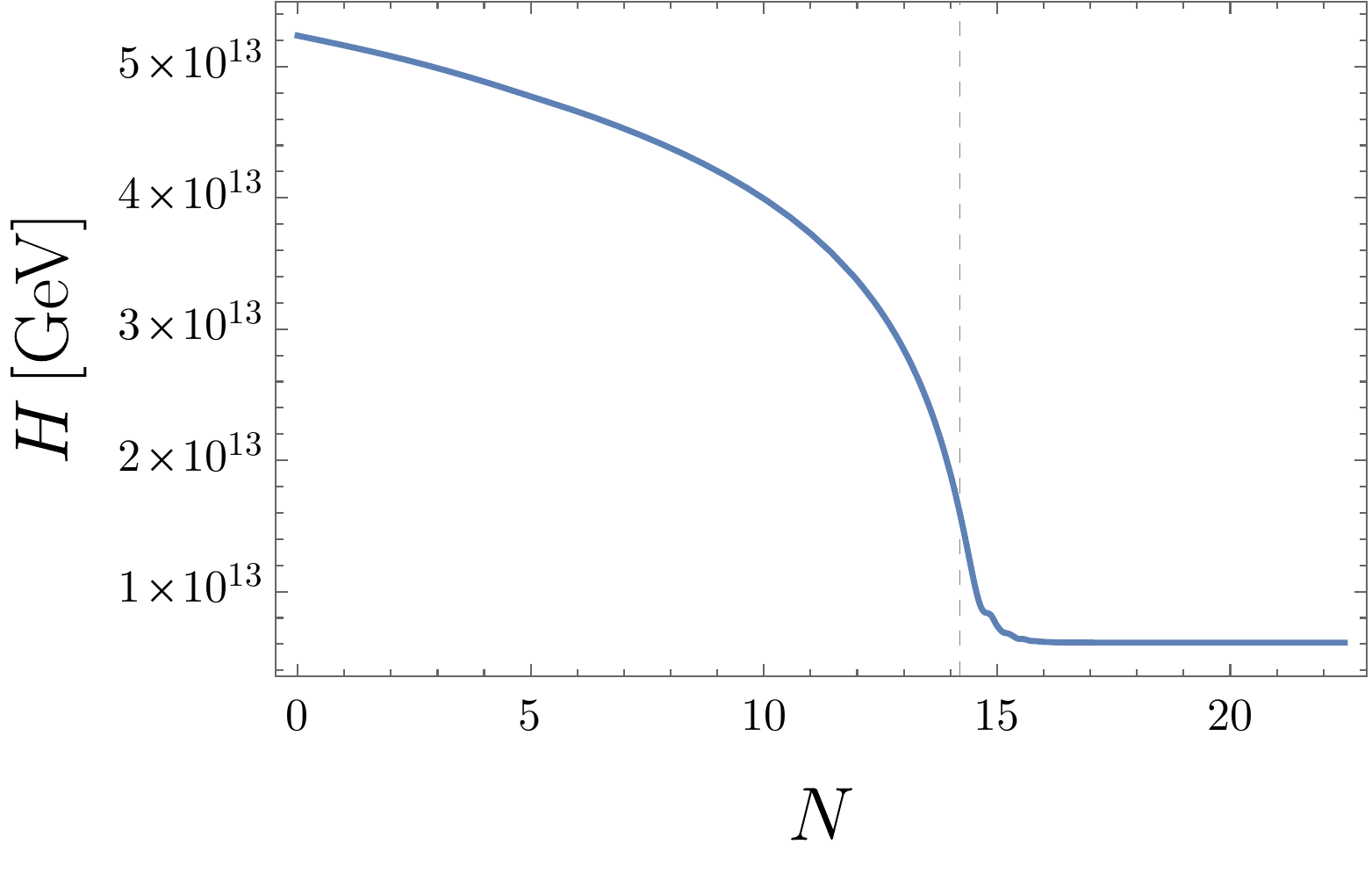}}
    \caption{
        Time evolution of $\chi$~(upper left panel), $Q$~(upper right panel), $m_Q$, $\xi$~(lower left panel), and the Hubble parameter~(lower right panel).
        The horizontal axis is the e-foldings and $N=0$ is set to be the onset of the second stage of inflation.
        The vertical dashed lines represent the end of the second inflationary stage when $\epsilon=1$.
        The gray dashed line in the upper right panel represents the minimum of the potential $Q_{+}$~\eqref{Q_local}.
        We set~$ ( p, \, f, \, \lambda, \, \mu, \, g, \,  \Delta N ) = ( 0.1, \, M_{ \mathrm{Pl} }, \, 25, \, 2.36 \times 10^{16}~\mathrm{GeV}, 1.0 \times 10^{-3}, \, 37 ) $ and obtain~$ \chi _{ \mathrm{cr} } = -4.13 M_{ \mathrm{Pl} } $,~$ \chi _{ \mathrm{CMB} } = -6.68 M_{ \mathrm{Pl} } $, and~$ N_{ \mathrm{2nd} } = 14.2$.
        }
    \label{evolution_of_background_parameterset1} 
\end{figure}

Fig.~\ref{evolution_of_background_parameterset1} shows the time evolution of~$\chi$, $Q$, $m_Q$, $\xi$, and $H$ during the second and third stages. We fix the parameters as $\mu=2.36 \times 10^{16}~\mathrm{GeV}$ and $\Delta N=37$ in addition to Eq.~\eqref{parameter values}.
From the upper left panel of Fig.~\ref{evolution_of_background_parameterset1}, one can see that, during (after) the second stage of inflation, the axion field slowly rolls the potential (quickly oscillates with a decaying amplitude).
In the upper right panel, the blue and gray lines represent the evolution of~$Q$ and the value of the true vacuum~$Q_{+}$ calculated based on Eq.~\eqref{Q_local}.
One observes that $Q$ follows $Q_+$ with a slight delay and catches up at $N\simeq 5$. They move together during the rest of the second inflationary stage by keeping the CN system.
Right after the end of the second inflation, $Q$ exhibits damped oscillations and quickly disappears.
The lower left panel indicates that $m_Q$ and $\xi$ maintain the relation~\eqref{ximQ} after $Q$ settles down to $Q_+$ at $N\simeq 5$ and reach their maximum value $\approx 15$ before exhibiting damped oscillations.
We can also see that $\xi$ is always larger than $\xi_\mathrm{cr}$ after $Q$ settles down to $Q_+$.
Therefore, the gauge fields do not experience the instability at $Q = Q_+$ existing only for $\xi < \xi_\mathrm{cr}$~\cite{Dimastrogiovanni:2012ew}.
The lower right panel presents that $H$ slowly decreases during the second stage, while it becomes constant in the third stage.

In Fig.~\ref{evolution_of_background_parameterset2}, we also present the time evolution of the energy densities and the slow-roll parameters of each component in our model.
From the left panel, we can see that during and after the second stage of inflation, the axion and additional fields dominate the universe, respectively.
From the right panel, we can see that the slow-roll parameter of the axion is the biggest for most of the time during the second stage, while $\epsilon_{ \mathrm{B} }$ also significantly contributes at around the transition to the third stage.
The total $\epsilon$ goes beyond unity for a short duration of time, and hence inflation pauses before resuming its third stage.

\subsubsection{The third stage: additional inflation}

Finally, we discuss the third stage of inflation driven by the additional inflaton field.
For simplicity, we consider its energy density $\rho _{ \mathrm{add} }$ as a constant value and it dominates the universe once the axion and the gauge fields decay after the end of the second stage. 
The energy density of the additional field is determined as follows.
First, we numerically solve the background dynamics of the CN system without the additional field and evaluate the total energy density at the end of inflation.
Then, we add the additional field whose energy density is 10\% of that value and solve the dynamics again.
As a result, $\rho _{ \mathrm{add} }$ becomes $\mathcal{O} (10)$\% at the end of the second inflation in the total system.
We confirmed that this size of $\rho _{ \mathrm{add} }$ does not significantly alter the dynamics during the second stage of inflation. 
We fix the duration of the third stage such that the total e-foldings of inflation is 50.

In this section, we have explained the background dynamics in our model. As seen in Fig.~\ref{evolution_of_background_parameterset1}, the background gauge fields have a large amplitude at the end of the second stage. Since it is known that the amplitude of the sourced GWs is exponentially sensitive to $m_Q$, we expect the copious production of GWs. We will compute the sourced GWs in Sec.~\ref{Tensor Perturbation}.

\section{Tensor Perturbation}
\label{Tensor Perturbation}

In this section, we perform numerical simulations to investigate the tensor perturbations in our model.
In Sec.~\ref{EoM of Tensor Perturbation}, we explain the features of the tensor perturbations in the CNI and our model, and solve their equations of motion numerically.
In Sec.~\ref{Tensor Power Spectrum}, we calculate the primordial tensor power spectrum sourced by the CN system to estimate the GW spectrum today.

\subsection{EoM for Tensor Perturbation}
\label{EoM of Tensor Perturbation}

The spacetime metric with tensor perturbations is defined as
\begin{equation}
    \mathrm{d} s ^2
    =
    g_{ \mu \nu } \mathrm{d} x^{\mu} \mathrm{d} x^{\nu} 
    =
    a^2 \, \left[ - \mathrm{d} \tau ^2 + ( \delta _{ i j } + h _{ i j } ) \mathrm{d} x^i \mathrm{d} x^j \right],
\end{equation}
where~$ \tau $ is the conformal time, which is related to the physical time by the relation~$ a \mathrm{d} \tau = \mathrm{d} t $.
With the nonzero gauge field background in the CN system, the SU(2) gauge symmetry and the spatial SO(3) symmetry are spontaneously broken into a diagonal SO(3).
Then, the gauge index of the SU(2) gauge fields can be identified as the spatial index.
The perturbations of the SU$(2)$ gauge fields are decomposed as
\begin{equation}
    \delta A^a _i
    =
    t _{ a i } + \ldots
\end{equation}
where $t_{ai}$ is the tensor perturbation, and $\ldots$ denotes the scalar and vector perturbations, which are decoupled from the tensor components and hence we neglect in this paper.

We impose the transverse-traceless condition on~$ h_{ i j } $ and~$ t_{ i j } $.
Let~$ h_{+}, \, h_{\times} $ ($ t_{+}, \, t_{\times} $) be the plus and cross mode for metric (gauge field) tensor perturbations.
We define the right~(R) and left~(L)-handed modes of the tensor perturbations as
\begin{equation}
    t_{ \mathrm{R/L} }
    =
    t_{+} \mp i t_{\times},
    \qquad
    \psi _{ \mathrm{R/L} }
    =
    \frac{ a M_{ \mathrm{Pl} } }{2} \, ( h_{+} \mp i h_{\times} ),
\end{equation}
where minus/plus sign represents R/L respectively.
Then, the equations of motion for~$ t_{ \mathrm{R/L} } $ and~$ \psi _{ \mathrm{R/L} } $ read
\begin{eqnarray}
    &&
    \partial _x ^2 t_{ \mathrm{R/L} } + \left[ 1 \mp \frac{ 2 ( m_Q + \xi ) }{ k/aH } + \frac{ 2 m_Q \xi }{ \left( k/aH \right)^2 } \right] t_{ \mathrm{R/L} }
    \nonumber
    \\
    &&
    =
    - \frac{ 2 \kappa _{ \mathrm{E} } }{ k/aH } \, \partial _x \psi _{ \mathrm{R/L} }
    +
    \frac{2}{ \left( k/aH \right)^2 } \, \left[ \left( m_Q \mp \frac{k}{ a H } \right) \, \kappa _{ \mathrm{B} } + ( 1 - \epsilon ) \, \kappa _{ \mathrm{E} } + \epsilon _{ \mathrm{E} } \, \eta _{ \mathrm{E} } \right] \psi _{ \mathrm{R/L} },
    \label{monodromygaugetensorEoM}
    \\
    &&
    \partial _x ^2 \psi _{ \mathrm{R/L} } + \left[ 1 - \frac{ 2 - \epsilon }{ \left( k/aH \right)^2 } \right] \psi _{ \mathrm{R/L} }
    =
    \frac{ 2 \kappa _{ \mathrm{E} } }{ k/aH } \, \partial _x t_{ \mathrm{R/L} } + \frac{ 2 \kappa _{ \mathrm{B} } }{ \left( k/aH \right)^2 } \, \left( m_Q \mp \frac{k}{ a H } \right) \, t_{ \mathrm{R/L} },
    \label{monodromygravitytensorEoM}
\end{eqnarray}
where $x \equiv -k \tau$,
$ \kappa _{ \mathrm{E} } = ( H Q + \dot{Q} ) / ( H M_{ \mathrm{Pl} } ) $, and
$ \kappa _{ \mathrm{B} } = g Q^2 / ( H M_{ \mathrm{Pl} } ) $
satisfying~$ \kappa _{ \mathrm{E} } ^2 = \epsilon _{ \mathrm{E} } $, $ \kappa _{ \mathrm{B} } ^2 = \epsilon _{ \mathrm{B} } $, and~$ \eta _{ \mathrm{E} } = \dot{ \epsilon } _{ \mathrm{E} } / ( H \epsilon _{ \mathrm{E} } ) $.
We note that the quantities~$H$,~$m_Q$,~$\xi$,~$\kappa _{ \mathrm{E} }$,~$\kappa _{ \mathrm{B} }$, and~$\eta _{ \mathrm{E} }$ generally depend on time, which are obtained by solving Eqs.~\eqref{Background_EoM_axion}, \eqref{Background_EoM_gauge_field}, and~\eqref{Friedmann_eq_our_model}.
$\kappa_{ \mathrm{E} }$ can be negative at the end of the second stage (see the upper right panel of Fig.~\ref{evolution_of_background_parameterset1}).

As can be seen in the left-hand side of Eq.~\eqref{monodromygaugetensorEoM}, only~$t_{ \mathrm{R} }$ undergoes the tachyonic instability, not~$t_{ \mathrm{L} }$ due to the positivity of~$m_Q$ and~$\xi$.
One can also see that $t_{ \mathrm{R} }$ sources only $\psi_{ \mathrm{R} }$ in Eq.~\eqref{monodromygravitytensorEoM}.
Thus, we concentrate on the right-handed tensor modes~$ \psi _{ \mathrm{R} } $ and $ t_{ \mathrm{R} } $.
We find the condition for the tachyonic instability as
\begin{equation}
    \frac{ 2 ( m_Q + \xi ) }{ k/aH }
    >
    \mathrm{max} \left\{ 1, \, \frac{ 2 m_Q \xi }{ \left( k/aH \right)^2 } \right\}
    \label{Condition_tachyonic_instability}
    \quad
    \Leftrightarrow\quad
    \, \, \, 
    \frac{ m_Q \xi }{ m_Q + \xi }
    <
    \frac{k}{ a H }
    <
    2 \, ( m_Q + \xi ) .
\end{equation}
This inequality implies that the tachyonic instability occurs before or around the horizon exit, since $ m_Q \xi / ( m_Q + \xi ) $ is~$ \mathcal{O} (1) $ or larger in the regime of interest.

Once the tachyonic instability disappears and the source term becomes negligible, the sourced tensor mode of the metric perturbations obeys the equation~$ \partial _x ^2 \psi _{ \mathrm{R} } + [ 1 - ( 2 - \epsilon ) \, ( a H / k )^2 ]  \psi _{ \mathrm{R} } = 0 $, which has the solution~$ \psi _{ \mathrm{R} } \propto a $ at the superhorizon limit~$ k \ll a H $.
Thus, $\psi_{ \mathrm{R} } / a$ conserves into a constant value in the superhorizon limit.
In the de Sitter spacetime, which is a good approximation during the third stage of inflation, $ x \, \psi _{ \mathrm{R} } $ is also conserved.

We numerically solve Eqs.~\eqref{monodromygaugetensorEoM} and \eqref{monodromygravitytensorEoM}, using the background solutions of Eqs.~\eqref{Background_EoM_axion}, \eqref{Background_EoM_gauge_field}, and~\eqref{Friedmann_eq_our_model} shown in Figs.~\ref{evolution_of_background_parameterset1} and \ref{evolution_of_background_parameterset2}.
The initial condition for $t_{ \mathrm{R} }$ is the asymptotic solution in the sub-horizon regime,
$t_{ \mathrm{R} } ( x \gg 1) =   ( 2 x )^{-i ( m_Q + \xi )} e^{ i x  }/\sqrt{ 2 k }$.
On the other hand, the initial condition for the metric tensor is
 $ \psi _{ \mathrm{R} } = \partial _{x} \psi _{ \mathrm{R} } = 0$,
because we are interested in the sourced GWs (i.e. the inhomogeneous solution of Eq.~\eqref{monodromygravitytensorEoM}).

\begin{figure}
    \subfigure[][$ k = 1.5 \times 10^{12} \, \mathrm{Mpc} ^{-1} $]{%
        \includegraphics[width=7.0cm]{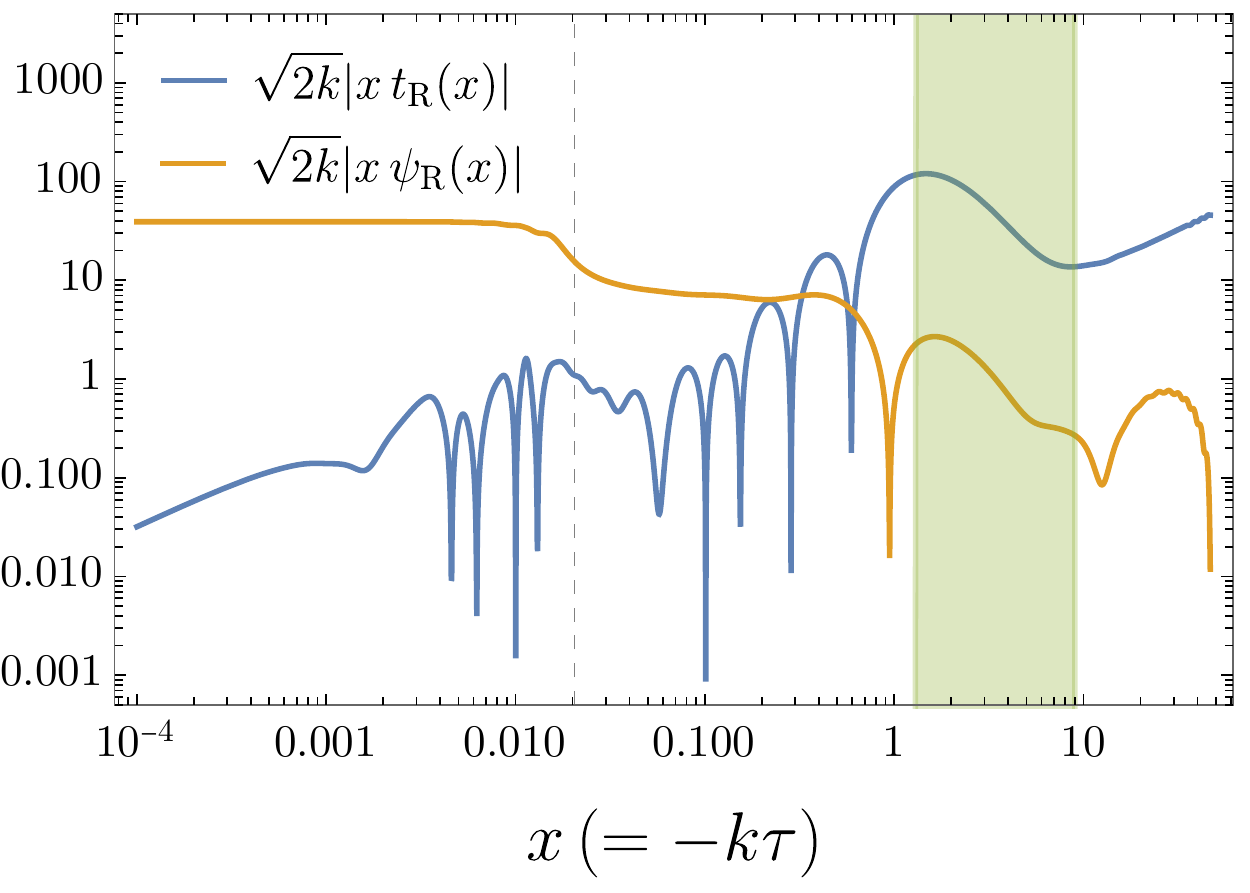}}%
    \subfigure[][$ k = 1.5 \times 10^{14} \, \mathrm{Mpc} ^{-1} $]{
        \includegraphics[width=7.0cm]{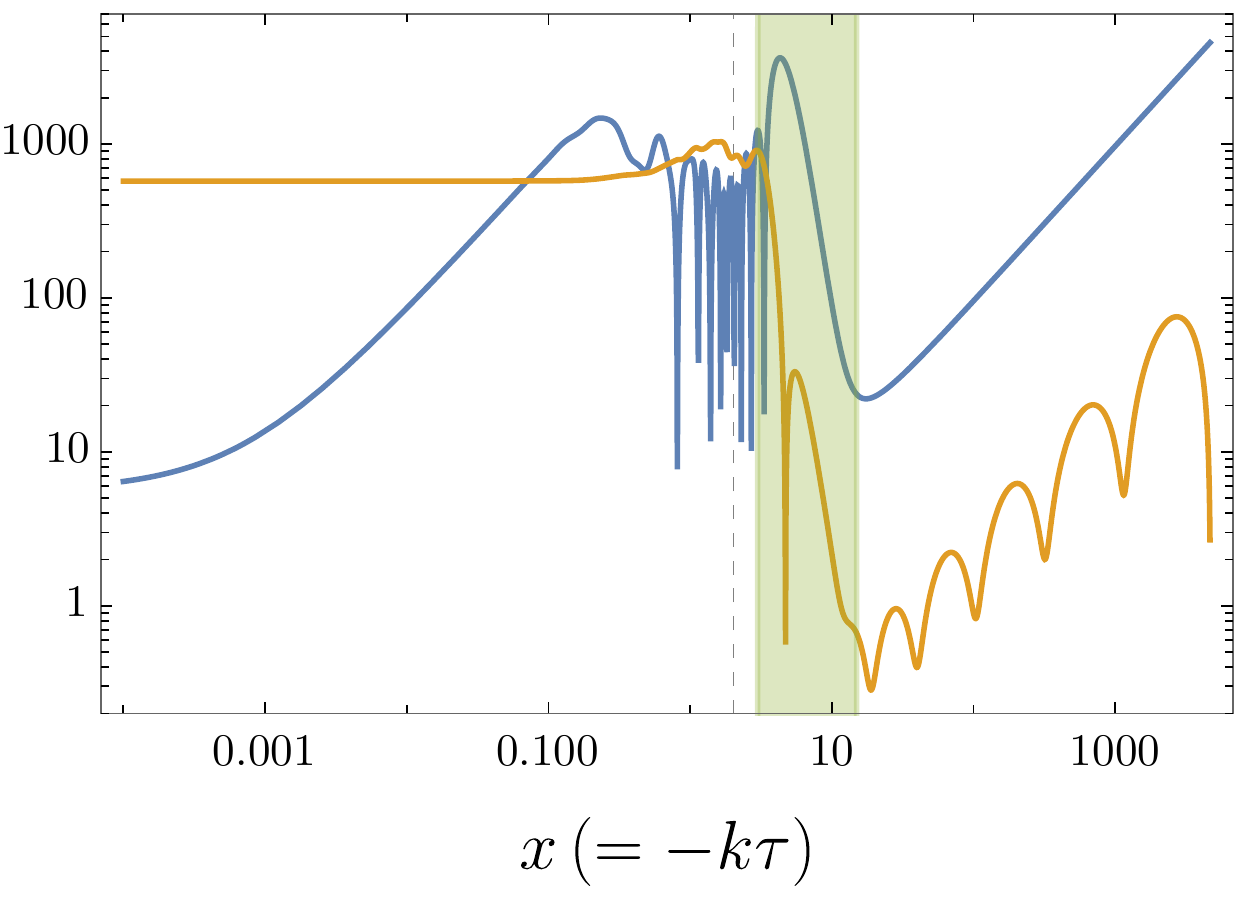}}
    \subfigure[][]{%
        \includegraphics[width=7.0cm]{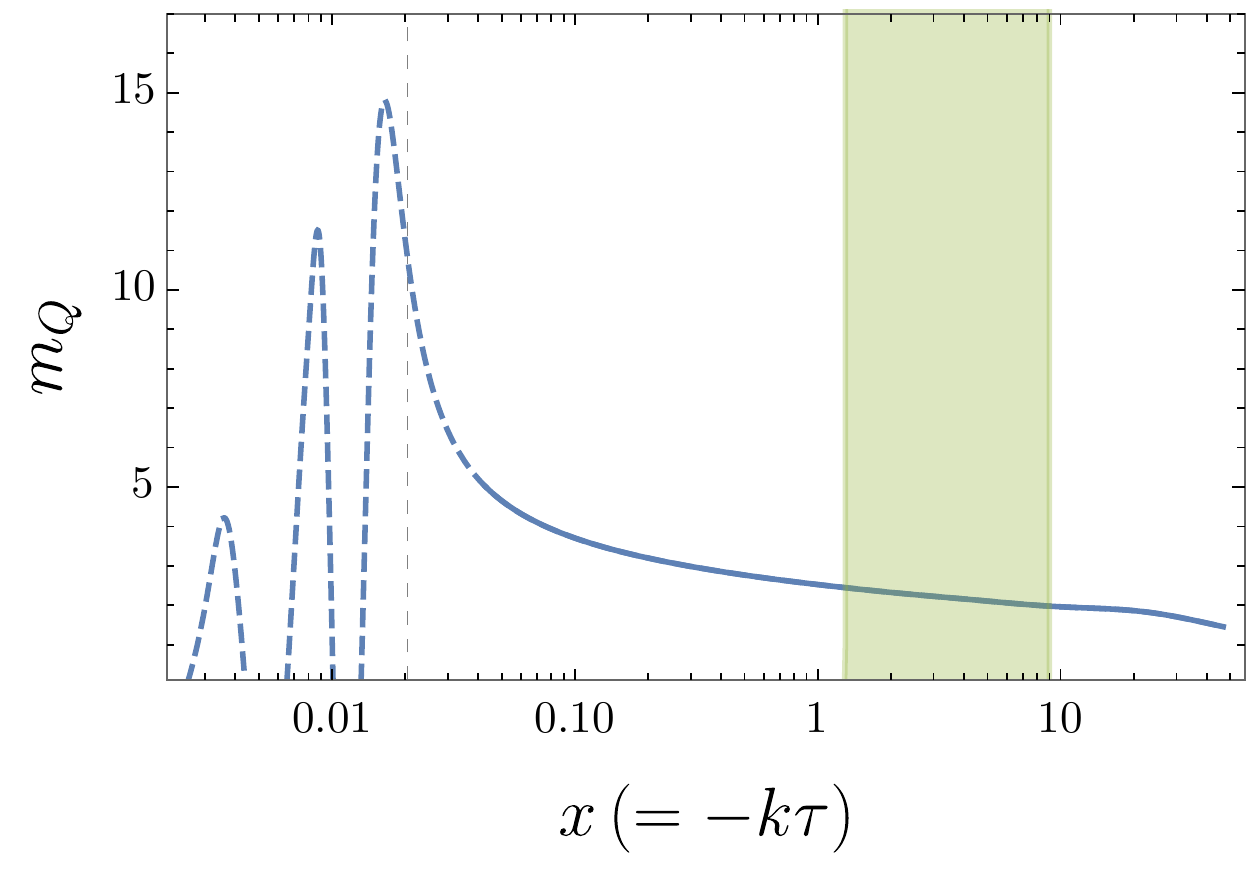}}%
    \subfigure[][]{
        \includegraphics[width=7.1cm]{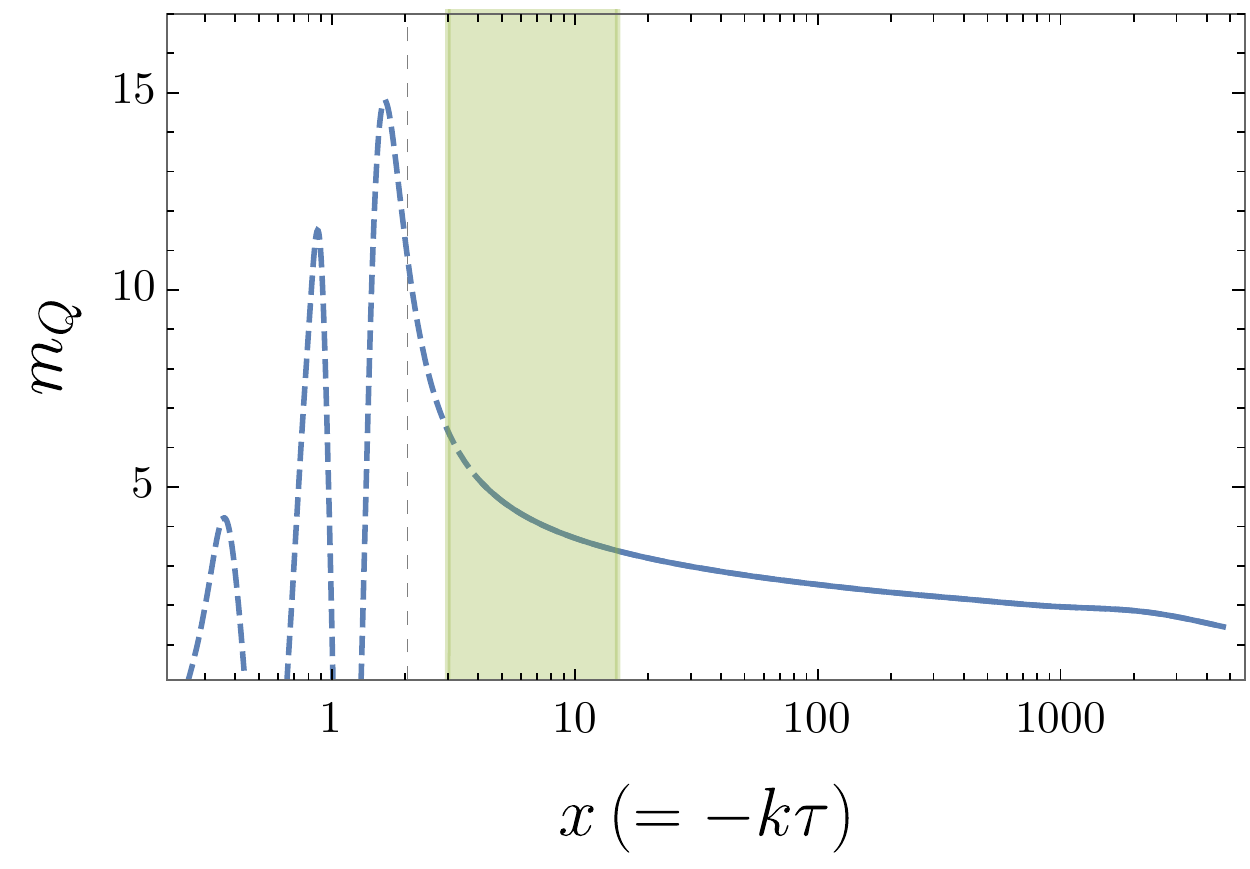}}
    \caption{
        \textbf{(a), (b)} Time evolution of tensor perturbations, $ \sqrt{2k} \, | x \, t_{ \mathrm{R} } ( x ) | $~(blue) and $ \sqrt{2k} \, | x \, \psi _{ \mathrm{R} } ( x ) | $~(orange), for the wavenumbers, $k=1.5\times 10^{12}~\mathrm{Mpc}^{-1}$ in (a) and $k=1.5\times 10^{14}~\mathrm{Mpc}^{-1}$ in (b).
        The green shaded regions represent the time interval when the condition for the tachyonic instability~\eqref{Condition_tachyonic_instability} is satisfied.
        The vertical gray lines denote the time when the second stage ends.
        \textbf{(c), (d)} Time evolution of~$ m_Q $, which is identical for (c) and (d) but overlapped with the instability regions of (a) and (b), respectively.
        After the condition for ignoring the backreaction~\eqref{Backreaction efficient} is violated, we show the time evolution of $m_Q$ with the dashed lines, because the backreaction of the tensor modes to the background dynamics is no longer negligible and our calculation is not reliable there.
        }
    \label{evolution_of_tensor_perturbation} 
\end{figure}
The panels of (a) and (b) in Fig.~\ref{evolution_of_tensor_perturbation} show the time evolution of~$ \sqrt{2k} \, | x \, t_{ \mathrm{R} } ( x ) | $~(blue) and $ \sqrt{2k} \, | x \, \psi _{ \mathrm{R} } ( x ) | $~(orange) with different wavenumbers.
To clarify when the tachyonic instability occurs, we also show the time interval satisfying the inequality~\eqref{Condition_tachyonic_instability} as the green shaded regions, which coincide with those in the panels (c) and (d) to illustrate the corresponding value of $m_Q$.
One can see that $ t_{ \mathrm{R} } $ is enhanced and sources $ \psi _{ \mathrm{R} } $ in these regions, and these effects are more significant for larger $m_Q$, as expected.
Therefore, for monotonically increasing $m_Q$, the larger the wavenumber is, the more significantly the tensor perturbations are enhanced.
One can also see that $ \sqrt{2k} \, | x \, \psi _{ \mathrm{R} } ( x ) | $ is conserved during the third stage of inflation.

We should stress that we have a limitation on $m_Q$ for which our analysis is consistent.
If the energy density of the enhanced tensor modes is no longer negligible compared to that of the background, the backreaction of the tensor modes to the background dynamics needs to be taken into account. The backreaction is not significant when~\cite{Fujita:2017jwq}
\begin{equation}
    e^{ 1.85 m_Q } g
    \lesssim
    10
    \, \, \, 
    \Leftrightarrow
    \, \, \, 
    m_Q
    \lesssim 
    m_{ Q, \mathrm{back} }
    \equiv
    \frac{1}{1.85} \, \ln \left( \frac{10}{g} \right),
    \label{Backreaction efficient}
\end{equation}
For the parameter used in this paper, $ g = 10^{-3} $, the threshold value is $ m_{ Q, \mathrm{back} } \simeq 4.98 $. In the region after this condition is first violated, $m_Q$ is denoted as the dashed lines in the bottom panels of Fig.~\ref{evolution_of_tensor_perturbation}.
Note that the backreaction from the tensor modes can also affect the background dynamics of the axion and then modify the duration of the second stage of inflation or $\Delta N$.
However, a modification of $\Delta N$ can be compensated by a change of the duration of the third stage of inflation. 
The perturbations that have already exited the horizon before the backreaction becomes significant are not affected.
As a result, the backreaction does not change the discussion on the CMB and GWs but only affects the interpretation of $\Delta N$.
If the backreaction considerably modifies the duration of the second stage of inflation, we need to interpret $\Delta N$ not as the duration of the first and second stages of inflation, but just as a free parameter that labels $(\chi_\mathrm{CMB}, \mu)$ consistent with $\mathcal{P}_\zeta$.

\subsection{Tensor Power Spectrum}
\label{Tensor Power Spectrum}

\begin{figure}[htpb]
        \begin{center}
            \includegraphics[clip,width=9cm]{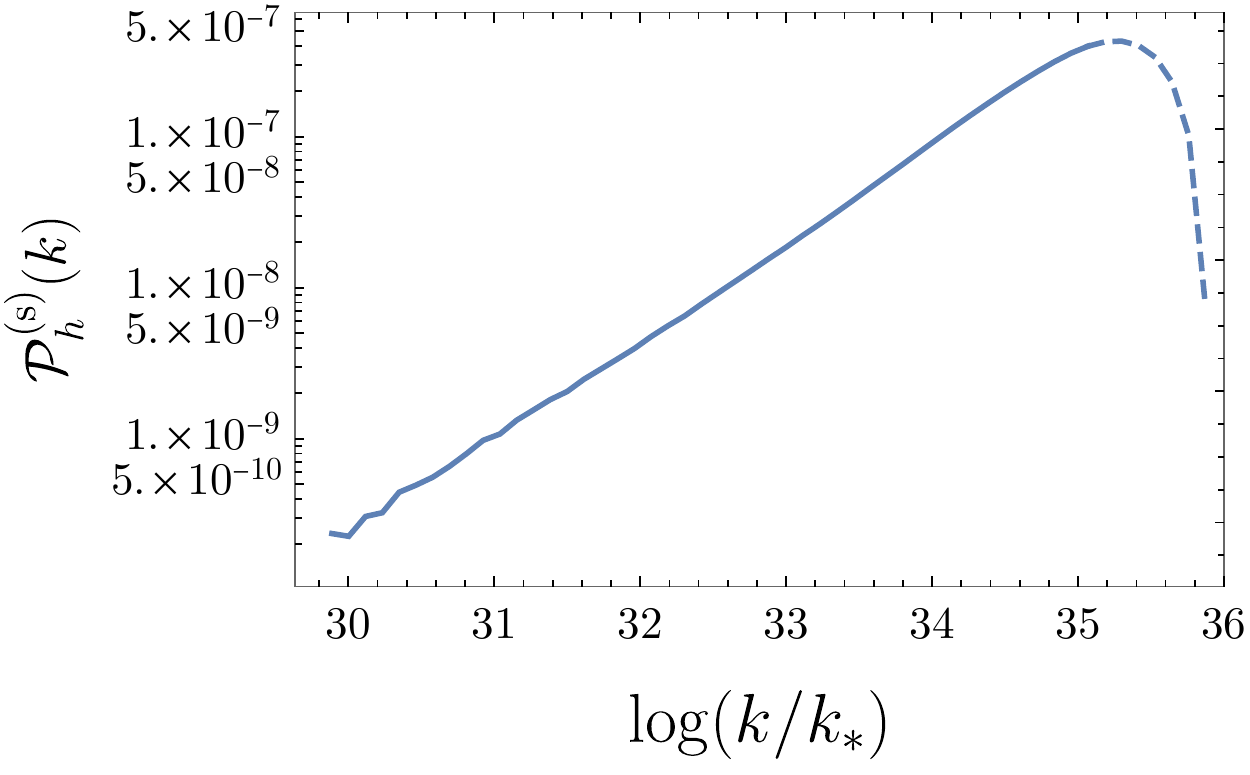}
            \caption{
                Primordial power spectrum of the sourced tensor perturbations evaluated with
                Eq.~\eqref{sourced_tensor_power_spectrum}.
                $k_{*} = 0.05 ~\mathrm{Mpc} ^{-1}$ denotes the CMB scale.
                The dashed blue line represents the region of Eq.~\eqref{Backreaction efficient} in which the backreaction of the tensor modes to the background dynamics is relevant.
                The background dynamics and the model parameters are the same as  in Figs.~\ref{evolution_of_background_parameterset1}--\ref{evolution_of_tensor_perturbation}.
                }
            \label{sourced_tensor_power_spectrum_fig}
        \end{center}
    \end{figure}

The primordial power spectrum of the sourced tensor perturbation in the superhorizon limit can be computed as
\begin{equation}
    \mathcal{P} ^{( \mathrm{s} )} _h (k)
    =
    \frac{ H^2 }{ \pi ^2 M_{ \mathrm{Pl} } ^2 } \, \left| \lim _{ x \rightarrow + 0 } \left[ \sqrt{ 2 k } x \, \psi ^{( \mathrm{s} )} _{ \mathrm{R} } (x) \right] \right|^2,
    \label{sourced_tensor_power_spectrum}
\end{equation}
where the Hubble parameter is evaluated at the third stage of inflation.
We evaluate $\mathcal{P} ^{( \mathrm{s} )} _h (k)$ with the numerically obtained $\sqrt{ 2 k } x \, \psi ^{( \mathrm{s} )} _{ \mathrm{R} }$.
The result is shown in Fig.~\ref{sourced_tensor_power_spectrum_fig}.
The power spectrum is blue-tilted for the aforementioned reason.
However, too large wavenumbers experience the tachyonic instability after the backreaction of the perturbations becomes relevant, which makes our estimations unreliable.
Thus, if the condition for ignoring the backreaction~\eqref{Backreaction efficient} is violated when a wavenumber satisfies the instability condition~\eqref{Condition_tachyonic_instability}, we consider the analysis of such a wavenumber to be invalid.
Such a wavenumber region is shown by the dashed line in Fig.~\ref{sourced_tensor_power_spectrum_fig}.

So far, we have not discussed the homogeneous solution of Eq.~\eqref{monodromygravitytensorEoM}, which corresponds to the vacuum fluctuations. 
The total power spectrum of the primordial tensor perturbation is given by
\begin{equation}
    \mathcal{P}_h (k)
    =
    \mathcal{P} ^{ (\mathrm{vac}) } _h (k) + \mathcal{P} ^{ (\mathrm{s}) } _h (k)
    =
    \frac{ 2 H^2 }{ \pi ^2 M_{ \mathrm{Pl} } ^2 } + \mathcal{P} ^{ (\mathrm{s}) } _h (k),
    \label{total_tensor_perturbation}
\end{equation}
where~$ \mathcal{P} ^{ (\mathrm{vac}) } _h (k) $ is the contribution of the vacuum fluctuation and the factor~$2$ in~$ \mathcal{P} ^{ (\mathrm{vac}) } _h (k) $ is due to the left and right polarization modes. 
However, for our choice of the model parameters, $\mathcal{P} ^{ (\mathrm{vac}) } _h$ is at most $\mathcal{O}(10^{-10})$ and is negligibly small compared to $\mathcal{P} ^{ (\mathrm{s}) } _h$ shown in Fig.~\ref{sourced_tensor_power_spectrum_fig}.

The GW spectrum today $\Omega _{ \mathrm{GW} }$ is obtained by solving the evolution equation for tensor modes from the end of the third stage of inflation until the present time.
The approximated expression can be derived as~\cite{Watanabe:2006qe}
\begin{equation}
    \Omega _{ \mathrm{GW} } ( k, \tau _0 )
    =
    \frac{ \mathcal{P} _h (k) }{ 12 H_0 ^2 } \, k^2 \cdot 
    \left\{ \,
        \begin{aligned}
            \frac{ \tau _{\mathrm{eq}} ^2 }{ \tau _0 ^2 } \left[ A(k) \, j_2 ( k \tau _0 ) + B(k) \, y_2 ( k \tau _0 ) \right] ^2 , \, \, \, \mathrm{if}
            \,\,
            k > k_{\mathrm{eq}},
            \\
            \left[ \frac{ 3 j_2 ( k \tau _0 ) }{ k \tau _0 } \right] ^2, \, \, \,
            \mathrm{if}
            \, \, 
            k < k_{\mathrm{eq}}, \hspace{3cm}
        \end{aligned}
    \right.
\end{equation}
with
\begin{eqnarray}
    A(k)
    &=&
    \frac{ 3 }{ 2 k \tau _{\mathrm{eq}} } - \frac{ \cos ( 2 k \tau _{\mathrm{eq}} ) }{ 2 k \tau _{\mathrm{eq}} } + \frac{ \sin ( 2 k \tau _{\mathrm{eq}} ) }{ ( 2 k \tau _{\mathrm{eq}} )^2 } ,
    \\
    B(k)
    &=&
    -1 + \frac{ 1 }{ ( k \tau _{\mathrm{eq}} )^2 } - \frac{ \cos (2 k \tau _{\mathrm{eq}}) }{ ( k \tau _{\mathrm{eq}} )^2 } - \frac{ \sin ( 2 k \tau _{\mathrm{eq}} ) }{ 2 k \tau _{\mathrm{eq}} } ,
\end{eqnarray}
where the lower index $0$ denotes the value of today, and $k_{\mathrm{eq}} \simeq 1 / \tau _{\mathrm{eq}} \simeq 2.38 \times 10 ^{-3}~\mathrm{Mpc} ^{-1}$ is the wavenumber of the modes that reenters the horizon at the matter-radiation equality.
In this formula, the transfer function oscillates rapidly with respect to the wavenumber.
Then we use the envelope of this formula for $ k > k_{\mathrm{eq}} $.
We also use the unit conversion~$ 1 \, \mathrm{Hz} = 6.5 \times 10^{14} \, \mathrm{Mpc} ^{-1} $.

\begin{figure}[htpb]
        \begin{center}
            \includegraphics[clip,width=14cm]{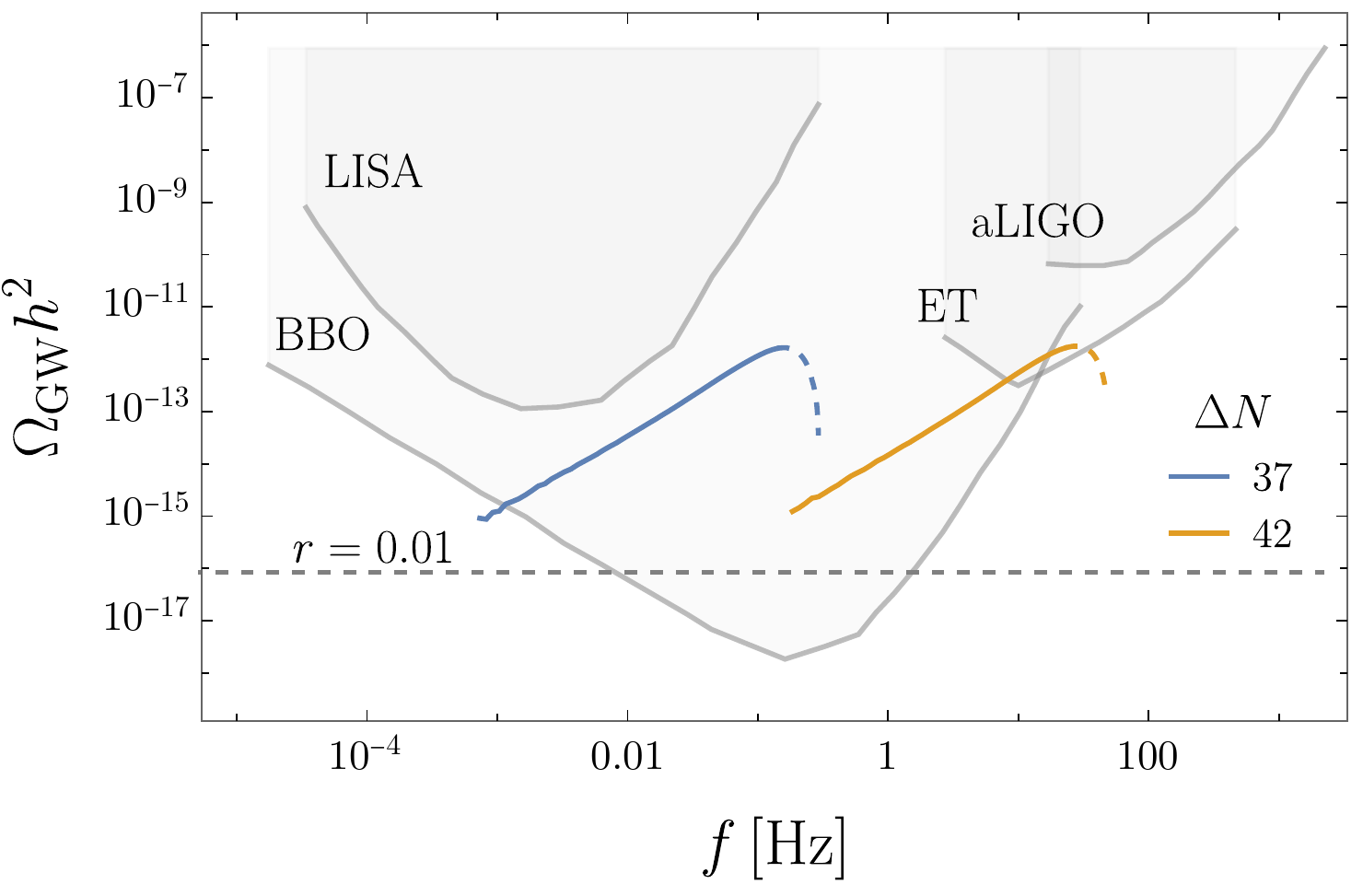}
            \caption{
                Gravitational waves today predicted in our model~(colored lines).
                The blue and orange lines represent the spectrum with $ \Delta N = 37$ and~$42$, respectively, and the dashed lines represent the range where the backreaction of the tensor modes to the background dynamics is not negligible and thus the prediction is not reliable.
                The gray solid lines represent the sensitivity curves of the various future GW experiments, LISA~\cite{amaroseoane2017laser}, BBO~\cite{Yagi:2011wg}, ET~\cite{Kallosh:2009xx}, and aLIGO~\cite{KAGRA:2021kbb}.
                The gray horizontal solid line represents the flat GW spectrum with the tensor-to-scalar ratio~$r=0.01$ for a reference.
            }
            \label{GW_today}
        \end{center}
    \end{figure}

The final results are shown in Fig.~\ref{GW_today}.
The colored lines are the predictions in our model, and the dashed lines represent the range where the backreaction of the tensor modes to the background dynamics is not negligible and thus the prediction is not reliable.
The gray solid lines represent the sensitivity curves of the various future GW experiments, LISA~\cite{amaroseoane2017laser}, BBO~\cite{Yagi:2011wg}, ET~\cite{Kallosh:2009xx}, and aLIGO~\cite{KAGRA:2021kbb}.
From Fig.~\ref{GW_today}, we can see that the prediction of GW in our model is much larger than the vacuum one, and whenever CNI ends, the shapes and amplitudes of the spectra are almost the same except for the peaked frequencies.
Although it is hard to recognize in Fig.~\ref{GW_today}, the peak height of the sourced GWs slightly shifts depending on $\Delta N$, because the parameter of the potential height $\mu$ changes accordingly, as discussed in Sec.~\ref{The first stage: single-field slow-roll phase}.

\section{Summary and Discussion}
\label{Summary and Discussion}

In this paper, we investigate the inflationary model which contains three stages: single-field slow-roll phase, CNI phase, and additional inflation to study the behavior of the tensor modes when CNI ends.
Since the minimal CNI is ruled out from the CMB observation due to the overproduction of GWs, we study the model in which the gauge fields have no VEV when the CMB scale exits the horizon, and after that, the gauge fields experience the phase transition obtaining nonzero VEVs.
In this model, the theoretical value of the tensor-to-scalar ratio at the CMB scale is smaller than that of the minimal CNI because the prediction in our model is the same as that of single-field inflation.
Moreover, we add another inflationary period caused by the additional field after CNI and thereby, the GW signals in our model become detectable by the various future GW observations.

The background dynamics in our model is as follows.
We consider the situation where the parameter~$\xi$ proportional to the velocity of the axion grows as the axion rolls down the potential.
With a small axion velocity, the effective potential of the background gauge fields has the true vacuum at the origin.
When the parameter~$\xi$ exceeds a critical value~$\xi _{ \mathrm{cr} }$, which implies that the axion rolls fast enough, the effective potential of the background gauge fields has two vacua: the trivial solution and the non-trivial solution with nonzero VEVs, and the latter becomes the true one.
We assume that the perturbations of the gauge fields have grown enough to exceed the potential barrier from the false vacuum to the true one when $\xi = \xi_\mathrm{cr}$.
When solving the CN system, we use the Chern--Simons coupling~$\lambda$ of~$ \mathcal{O} (10) $ for the realization of the phase transition
which is smaller than typical values in the minimal CNI.
In our numerical calculation solving the CN system, the initial field values of the background gauge fields are set to the local maximum of the effective potential as an approximation of the gauge fields overcoming the potential barrier.
Then, the gauge fields fall into the true vacuum, and after that, their VEVs gradually grow following the motion of the true vacuum.
We do not solve the equation of motion with the backreaction of the tensor perturbations to the background.
Thus, the prediction in our numerical simulation is not reliable after the VEVs of the gauge fields exceed a certain value where the strong backreaction of the tensor modes to the background dynamics becomes relevant.

The dynamics of the tensor perturbations in our model is as follows.
We solve the equations of motion for the tensor perturbations derived without the slow-roll approximation in order to investigate the whole period of the second stage including its end.
As the VEVs of the gauge fields grow in time, the enhancement of the tensor modes through the tachyonic instability also becomes more significant and the enhanced tensor modes are converged into larger values in the superhorizon limit.
The GWs today predicted in our model are detectable by various future GW observations such as BBO and ET.

We assumed that the fluctuations of the gauge fields grow enough to realize the transition of the background field at the onset of the second stage and used an ad hoc configuration as the initial condition for the second stage.
However, for a more precise discussion, we need to perform the classical lattice simulation or develop a dedicated treatment, which approximates the system by a simpler one, to investigate the behavior of the phase transition of the gauge fields.

\section*{Acknowledgments}

We would like to thank Takashi Hiramatsu and Ryo Namba for useful comments.
This work is supported by the Grant-in-Aid for Scientific Research Fund of the JSPS 18K13537 (T.\,F.) and 20J20248 (K.\,M.).
K.\,M. is supported by World Premier International Research Center Initiative (WPI Initiative), MEXT, Japan and the Program of Excellence in Photon Science.

\appendix

\section{The effect of the gauge fields during the first stage}
\label{The effect of the gauge fields during the first stage}

In Sec.~\ref{The first stage: single-field slow-roll phase}, we compute the inflaton dynamics by neglecting its coupling to the gauge fields during the first stage of inflation. 
Here we verify this single-field slow-roll approximation. When the SU(2) gauge fields do not have VEVs, they can be viewed as three copies of U(1) gauge fields, because their self-interactions are ineffective.
The effect of the coupling has been studied in previous works and the conditions to safely ignore it are given by~\cite{Barnaby:2011vw}
\begin{align}
\mathcal{P} _{\mathcal{R}} \gg
3\mathcal{P}_{\mathcal{R}}^{\rm 1loop}
&\qquad\Leftrightarrow\qquad
3\mathcal{P} _{\mathcal{R}}\,f_2(\xi) e^{4\pi\xi}
\ll 1,
\label{1loop power spectrum}
\\
|3H\dot{\chi}|\simeq |V'(\chi)| \gg
\frac{3\lambda}{f}|\langle \vec{E}\cdot\vec{B}\rangle|
&\qquad\Leftrightarrow\qquad
\frac{H^2}{2\pi|\dot{\chi}|}\,\ll\,
7\xi^{3/2}e^{-\pi\xi},
\label{BG inflaton unaffected}
\end{align}
where we multiplied the right hand side by a factor of three for the three copies of $U(1)$ gauge fields.
The first condition implies that the one-loop contribution $\mathcal{P}_{\mathcal{R}}^{\rm 1loop}$ from the vector perturbation to the curvature power spectrum is negligible, while the second condition ensures that the effect on the background inflaton dynamics is negligible. The full shape of $f_2(\xi)$ can be found in Ref.~\cite{Barnaby:2011vw}.

When the CMB modes exit the horizon, namely at $\chi=\chi_\mathrm{CMB}$, the value of $\xi$ in our calculation is $\xi_\mathrm{CMB}\approx 1$. Since $f_2(1)\simeq 2\times 10^{-5}$, 
the left hand side of Eq.~\eqref{1loop power spectrum} reads $3\mathcal{P} _{\mathcal{R}}\,f_2(\xi) e^{4\pi\xi}\simeq 4\times 10^{-8}$. Thus, the one-loop correction to the curvature power spectrum is completely negligible on the CMB scale. Even for the critical value $\xi_\mathrm{cr}\approx 2.12$,
one finds $3\mathcal{P} _{\mathcal{R}}\,f_2(\xi) e^{4\pi\xi}\simeq 1.3 \times 10^{-3}$ for $\mathcal{P} _{\mathcal{R}}=2.2\times 10^{-9}$ and the first condition is satisfied. Next, we consider the second condition.
By using the slow-roll approximation $H^2/(2\pi|\dot{\chi}|)\simeq \mathcal{P} _{\mathcal{R}}^{1/2}\approx 4.7\times 10^{-5}$, one can find the value of $\xi$ which 
saturates Eq.~\eqref{BG inflaton unaffected}
as $\xi\approx 4.5$. During the first stage of inflation, $\xi$ is always smaller than the critical value $\xi_\mathrm{cr}\approx 2.12$. Thus, the gauge fields do not significantly affect the background inflaton dynamics.

\small
\bibliographystyle{JHEP}
\bibliography{bibtex}

\end{document}